\documentclass[twocolumn,trackchanges]{aastex61}
\usepackage{gensymb}
\usepackage{rotating,epsfig}
\usepackage{natbib}
\usepackage{amsmath,amstext}
\usepackage[T1]{fontenc}
\usepackage{apjfonts} 
\usepackage[figure,figure*]{hypcap}

\def\msun{$\rm M_{\sun}$}
\def\rsun{$\rm R_{\sun}$}
\def\lsun{$\rm L_{\sun}$}
\def\teff{$\rm T_{eff}$}
\def\Rc{$\rm R_{C}$}
\def\Ic{$\rm I_{C}$}
\def\av{$\rm A_V$}
\def\degree{$^{\circ}$}
\def\SOri{${\sigma}$ Orionis\space}
\def\ewha{EW[H$\alpha$]}

\def\msunyr{M_{\sun} \, yr^{-1}}

\shorttitle{An IMTTS with a transitional disk in Orion OB1}
\shortauthors{P\'{e}rez et al.}

\received{2018}
\revised{2018}
\accepted{}
\begin{document}

%\title{An intermediate mass T Tauri star surrounded by a transitional disk in the sparse population of the Orion OB1 association}
\title{
%{\bf toward the \SOri cluster}}
A
transitional disk around an intermediate mass star
in the sparse population of the Orion OB1 association}
%{\bf !!!! Favor ver abajo}}
%The most massive stars surrounded by a transitional accreting disk in the \SOri stellar cluster}

\author{Alice P\'erez-Blanco}
\affil{University of Leeds,School of Physics and Astronomy, LS29JT, Leeds, UK.}

\author{Karina Mauc\'{o}}
\affiliation{Instituto de Radioastronom\'{\i}a y Astrof\'{\i}sica (IRyA), Universidad Nacional Aut\'{o}noma de M\'{e}xico, Morelia, Mexico}

\author{Jes\'us Hern\'andez}
\affiliation{Instituto de Astronom\'ia, Universidad Nacional Aut\'onoma de M\'exico, Ensenada, BC, Mexico}

\author{Nuria Calvet}
\affiliation{Department of Astronomy, University of Michigan, 500 Church Street, Ann Arbor, MI 48109, USA}

\author{Catherine Espaillat}
\affiliation{Department of Astronomy, Boston University, 725 Commonwealth Avenue, Boston, MA 02215, USA}

\author{Melissa McClure}
\affiliation{Anton Pannekoek Institute for Astronomy, University of Amsterdam, Science Park 904, 1098 XH Amsterdam, The Netherlands}

%\affiliation{European Southern Observatory, Karl-Schwarzchild-Str. 2, D-85748, Garching bie M\"{u}nchen, Germany}

\author{Cesar Brice\~{n}o}
\affiliation{Cerro Tololo Interamerican Observatory, Casilla 603, la Serena, Chile}

\author{Connor Robinson}
\affiliation{Department of Astronomy, Boston University, 725 Commonwealth Avenue, Boston, MA 02215, USA}

\author{Daniel Feldman}
\affiliation{Department of Astronomy, Boston University, 725 Commonwealth Avenue, Boston, MA 02215, USA}

\author{Luis Villarreal}
\affiliation{Postgrado de F\'{\i}sica Fundamental, Universidad de Los Andes, M\'{e}rida, Venezuela}

\author{Paola D'Alessio}
\affiliation{Instituto de Radioastronom\'{\i}a y Astrof\'{\i}sica (IRyA), Universidad Nacional Aut\'{o}noma de M\'{e}xico, Morelia, Mexico}

\begin{abstract}
We present a detailed study of the disk around the intermediate mass star SO~411, 
aiming to explain the spectral energy distribution of this star. We show that this 
is a transitional disk truncated at $\sim$11 au, with $\sim$0.03 lunar masses of 
optically thin dust inside the cavity. Gas also flows through the cavity, since we 
find that the disk is still accreting mass onto the star, at a rate of $\sim 5 \times 10^{-9}\msunyr$. 
Until now, SO~411 has been thought to belong to the $\sim$3 Myr old \SOri cluster. 
However, we analyzed the second Gaia Data Release in combination with kinematic data previously 
reported, and found that SO~411 can be associated with an sparse stellar population 
located in front of the \SOri cluster. If this is the case, then SO~411 is older and 
even more peculiar, {since primordial disks in this stellar mass range are scarce for 
ages $>$5 Myr}. Analysis of the silicate 10{\micron} feature of SO~411 indicates that 
the observed feature arises at the edge of the outer disk, and displays a very high 
crystallinity ratio of $\sim$0.5, with forsterite the most abundant silicate crystal. 
The high forsterite abundance points to crystal formation in non-equilibrium conditions. 
The PAH spectrum of SO~411 is consistent with this intermediate state between the hot and 
luminous Herbig Ae and the less massive and cooler T Tauri stars. Analysis of the 7.7{\micron} 
PAH feature indicates that small PAHs still remain in the SO~411 disk.
\end{abstract}


\section{Introduction}

Circumstellar disks, a natural byproduct of the star forming process, exhibit a variety of characteristics that have been attributed to the disks being in different evolutionary 
phases. They start being gas-rich,  which we call ``primordial'', and optically  thick, and evolve into dusty debris disks. Planetary systems may form in the disk along this evolution.
\citep[e.g.] []{calvet2011,williams2011,testi2014,alexander2014}.

Ground based and specially  {\it Spitzer} observations revealed 
crucial evolutionary phases in disk evolution. These phases were
represented by the transitional disks \citep[TD;][]{strom1989, calvet2005} 
and the pre-transitional disks \citep[PTD;][]{espaillat2007a,espaillat2007b}. 
TDs were characterized observationally as having small or no flux excess above the photosphere in the near infrared (NIR; $\lesssim$10$\mu m$) and excesses consistent with primordial, optically thick disks at longer wavelengths \citep{dalessio2009,espaillat2014}. 
PTDs had spectral energy distributions (SEDs) similar to those of TDs, except for a flux excess consistent with optically thick material at the shorter NIR wavelengths. 
Large holes (in TDs) or wide gaps (in PTDs) in the inner dust distribution 
of protoplanetary disks could explain the deficit of NIR excesses in the SED,
and this was consistent with the cavities observed by millimeter interferometric observations
\citep{andrews2011,casassus2016}. The much higher spatial resolution observations provided by the Atacama Large Millimeter and Submillimeter Array (ALMA) have revealed
that disks are highly structured,  with multiple gaps, arms, and other structures, that may or may not be apparent in the SED \citep{alma2015, andrews2016}.

Dynamical clearing by planets \citep{zhu2011,zhu2012, espaillat2014, owen2016,ruiz2016},
photoevaporation, dust grain growth, differential dust drift, dead zones, condensations fronts, have been invoked to explain the dust structures seen in protoplanetary disks 
\citep[e.g.][]{alexander2014,chiang2007,dullemond2005,birnstiel2012,pinilla2016,zhang2015}. Nonetheless, the large, $\sim$ tens of au, cavities in small and large grains present in TDs and PTDs make them stand out as  clear cases of disks that have  undergone significant structural changes since they formed.

The TD/PTD phase in  the low mass, Classical T Tauri stars
(CTTS) represents one of the evolutionary pathways from primordial to second generation disks \citep[e.g.][]{currie2009,hernandez2010,cieza2013}. An alternative evolutionary pathway is represented by the evolved disks \citep{hernandez2010,muzerolle2010,ercolano2017}, which exhibit weaker infrared excesses than the median SED of Taurus at all wavelengths
\citep{furlan2006}. {The evolved disks are also named "anemic disks", "homologously depleted disks", "weak-excess disks" or "dust depleted disks" \citep{lada2006,currie2009,muzerolle2010,sicilia2013}}.

It is not clear if there are comparable evolutionary paths for the higher mass Herbig Ae/Be stars (HAeBe). \citet{meeus2001} classified the SEDs of HAeBe into group I and group II, 
with group I showing stronger mid and far infrared excesses than group II. 
In principle, these two classes could be identified with the two evolutionary pathways in CTTS. However, \citet{dullemond2004} suggested an evolutionary scheme from group I (flared disks) to group II (flat disks), which is supported by the higher amount of small dust grains in group I found by Herschel/PACS photometry \citep{pascual2016}.
An argument against this evolutionary interpretation was the discovery using mid-IR imaging that group I objects had cavities, while they were not found in group II \citep{maaskant2013}. On the other hand, the recent discovery of cavities in large grains
in the disks of group II sources \citep{zhang2016,rubinstein2018}
adds another uncertainty to these alternatives.
Therefore, additional studies are necessary to reveal whether group I and group II form an evolutionary sequence or if these two groups follow different evolutionary pathways as those suggested for the CTTS mass range. 

Intermediate Mass T Tauri Stars \citep[IMTTS;][]{calvet2004}, showing emission lines and spectral types ranging from late F to early K, represent the link between  the HAeBe (with spectral types B, A or early F) and the CTTS (with spectral types K or M). Since IMTTS evolve along radiative tracks, they will eventually become HAeBe stars, Vega type stars, or hybrid disks around A type stars such as those reported by \citet{pericaut2017}. This last class of objects, which have gas to dust ratio enhanced  by two or three orders of magnitudes compared to primordial disks, could represent the last stage between the final 
phases of the primordial component in the disks and the beginning of the debris disk phase \citep{pericaut2017,espaillat2017}.

Detailed studies of significant samples of IMTSS in different phases of evolution
could improve our understanding of the evolutionary process connecting the two ranges of stellar masses represented by the CTTS and the HAeBe stars.
However, collecting these samples is difficult.
The number ratio of transitional disks to primordial disks around 
CTTS suggests a timescale of transitional disks clearing of 10\%-20\% of the lifetime of primordial disks \citep{muzerolle2010,luhman2010}.  
Since the disk dispersal mechanisms are more efficient at higher stellar masses 
\citep{hernandez2007,hernandez2009,lada2006,sung2009,ribas2015}, 
it is not surprising that less than 10\% of the known TD/PTD are in the IMTTS regime and only $\sim$2\% are  associated to late F type stars \citep{espaillat2014,vandermarel2016}.
In this paper we present a detailed study of one of the few transitional
disks known to be associated with a 5 to 10 Myr F type star.

SO~411 (also named HD 294268, BD-02 1321, PDS 119) is a F type IMTTS 
believed until now to belong to the \SOri cluster \citep{torres1995,caballero2008,hernandez2014}. 
However, the general region of the \SOri cluster can include stars from a foreground pre-main sequence stellar population kinematically separated by $\sim$7 kms$^{-1}$ in radial velocity
\citep{jeffries2006,maxted2008,hernandez2014}. This foreground stellar population could have a median age and distance similar to older stellar groups associated with
the sparser Orion OB1a subassociation \citep{briceno2007,briceno2018}.
Here, we present a spatial and kinematical analysis
based on Gaia data that indicates that SO~411 most likely belongs to this foreground
population.

The SED constructed from Spitzer photometry (IRAC and MIPS) indicates that SO~411 is surrounded by a TD/PTD disk \citep{hernandez2007}. 
A width of H$\alpha$ at 10\% of the line peak of $\sim$294 km/s, measured from high-resolution spectra \citep[R$\sim$34000;][]{hernandez2014}, indicates that the SO~411 disk is
accreting mass onto the star \citep{white2003}. Finally, its location in the region of PAH emission-dominated sources in IRAC color-color diagrams \citep{hernandez2007,gutermuth2009} suggests that SO~411 has PAH emission features. Visual inspection of the Spitzer IRS spectrum confirms the presence of PAH emission features and also shows silicate emission at 10{\micron} indicating the presence of sub-{\micron} grains close to the star 
\citep{sargent2009b}. Thus, SO~411 represents an unique opportunity to study in detail an accreting TD/PTD around an IMTTS that shows evidence of small dust close to the star.

We present a detailed modeling effort
of SO~411  using irradiated disks models \citep{dalessio1998,dalessio1999,dalessio2001,dalessio2006,espaillat2010}, which allows us to study the dust structure and mineralogy of its disk. This paper is organized as follows:  Section \S\ref{sec:obs} describes the observational data. 
Based on the second release of Gaia's data \citep[GAIA-DR2;][]{GAIA2018}, we present the astrometric and kinematic properties of SO~411 and the \SOri cluster in \S\ref{sec:gaia}.
Additional properties of the target are analyzed in\S{\ref{sec:param}, a near infrared excess study in \S\ref{sec:spex}, 
and the disk modeling in \S\ref{sec:model}.
We then discuss the implications
of the dust composition
in \S\ref{sec:sil} and 
the presence of PAH features in \S\ref{sec:pah}. Finally, in Section \S{\ref{sec:summ} we present our summary and conclusions.

\section{Observations} \label{sec:obs}

\subsection{Photometric data}
We obtained optical photometry of SO~411 on February 2016 using the Large Monolithic Imager on the 4.3 meter 
Discovery Channel Telescope (DCT) at the Lowell Observatory (Arizona - USA). The images were reduced fo\-llo\-wing the standard IRAF 
procedure to perform bias subtraction and sky-flat field correction. The photometry was obtained and calibrated 
following the procedure in \citet{massey1992} using the Landolt standard field SA 98 \citep{landolt2009} observed 
at the same airmass as the target field ($\sim$1.35). Table \ref{tab:newphot} shows the optical photometric data of SO~411 in the Johnson-Cousin system. The photometric uncertainties are below 0.02 magnitudes.

\begin{deluxetable}{cccccc}%[b!]
\tablecaption{DCT optical photometry \label{tab:newphot}}
\tablecolumns{6}
%\tablenum{2}
\tablewidth{0pt}
\tablehead{
\colhead{Object} & \colhead{U} & \colhead{B} & \colhead{V} & \colhead{\Rc} & \colhead{\Ic} \\
\colhead{} & \colhead{mag} & \colhead{mag} & \colhead{mag} & \colhead{mag} & \colhead{mag}
}
\startdata
SO~411 &  11.07 & 10.96 &  10.50 & 10.08  & 9.84\\
\enddata
\end{deluxetable}

\subsection{IRS Spectra}

SO~411 was observed with the Infrared Spectrograph (IRS) on board the Spitzer Space Telescope \citep{Houck04} on 2007 October 11 as part of the program P30381 \citep{oliveira2006}. Observations were obtained in staring mode using the short-low (SL) resolution module which spans wavelengths from 5 to 14 {\micron} and the high resolution modules (SH and LH) which span wavelengths from 10 to 37 {\micron}. 
The SL spectrum was obtained from the Cornell Atlas of Spitzer/IRS Sources \citep[CASSIS;][]{cassis}. The reduction and extraction of the LH \& SH data was performed with optimal point source extraction option in the
Spectral Modeling, Analysis and Reduction Tool \citep[SMART;][]{smart2004}, which is an IDL-based processing and analysis tool for the IRS instrument.

\subsection{SpeX Spectra}

We obtained NIR spectra of SO~411 on 2015, February 6 using the micron medium-resolution spectrograph SpeX on the 3 meter NASA Infrared Telescope Facility (IRTF) at Mauna Kea Observatory. The spectra were obtained using the LXD long mode with a $0.5x15 \arcsec$ slit and a wavelength co\-ve\-ra\-ge of 1.9-4.6$\mu m$, giving a spectral resolution of R$\approx$1500, and using the lower resolution prism mode with a $3.0x15 \arcsec$ slit with wavelength coverage of 0.8-2.5$\mu m$, with R$\approx$75. The spectra were collected using the AB dither mode, in which the source is offset between two locations on the slit between the A and B exposure in order to subtract out sky emission and dark current for each AB spectra pair.

We reduced the observation using the 2014 version of the SpexTool, an IDL-based reduction package built s\-pe\-ci\-fi\-ca\-lly for the cross-dispersed spectra produced by SpeX \citep{cushing2004}. The reduction process consisted of flat-fielding, sky corrections and wavelength calibration using calibration frames taken during observations. Each order for every observation was extracted, stacked, corrected for telluric absorption features using the measured spectra of a standard A0V star and then stitched together to form a complete spectrum. Regions of low signal-to-noise were removed from the final spectra. A more complete description of the Spextool reduction process is described in \citet{cushing2004}. HD 34317 was the telluric A0V star used for the reduction  of the SpeX spectra.

\section{Analysis} \label{sec:ana}

\subsection{Spectral Type } \label{sec:SPT}

Here we revisited the optical spectrum from \citet{hernandez2014} to improve the spectral type of SO~411 obtained using 
the SPTCLASS tool, an IRAF/IDL code based on the methods described by \citet{hernandez2004}. The low resolution optical 
spectrum was obtained with the Boller \& Chivens spectrograph mounted on the 2.1 meter telescope at the San Pedro Martir Observatory. We used a 400 lines mm$^1$ dispersion grating along with a 2{\arcsec} slit width, giving a spectral resolution of $\sim$6{\AA} centered at $\sim$5500{\AA} \citep[for additional details, see ][]{hernandez2014}. 

Figure \ref{fig:sptclass} shows a comparison between the SO~411 spectrum and a group of standard spectra with similar resolution 
and spectral coverage obtained with  the FAST spectrograph \citep{fabricant1998} mounted on the 1.5 meter telescope of the Fred 
Lawrence Whipple Observatory and used to ca\-li\-bra\-te the SPTCLASS tool. 
In a previous effort we had estimated the spectral 
type of SO~411 as F7.5$\pm$2.5
\citep{hernandez2014}.
In order to reduce the uncertainty, we estimate a new spectral type using only the strongest spectral features that appear in IMTTS, such as the G-band($\lambda$ 4300\AA), \ion{Mn}{1}+\ion{Fe}{1}($\lambda$ 4458\AA), 
\ion{Fe}{1}($\lambda$ 5329\AA), \ion{Ca}{1}($\lambda$ 5589\AA), and \ion{Mn}{1}($\lambda$ 6015\AA). The improved spectral type of F6$\pm$1 is in agreement
with the previous estimate within the uncertainties.
 
\begin{figure}[t!]
\plotone{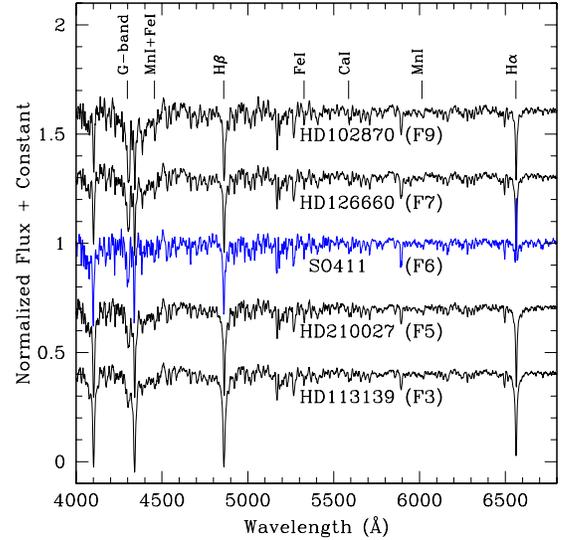}
\centering
\caption{Spectra of SO~411 compared to spectral type standards. Visual inspection and measured spectroscopic features indicate an spectral type of F6$\pm$1.\label{fig:sptclass}} 
\end{figure}

\subsection{Space-velocity analysis of SO~411} \label{sec:gaia}

Here we study the astrometric and kinematic properties of SO~411 and the \SOri cluster.
For a sample of known young stars defined in \citet{hernandez2014}, we combined the parallaxes and proper motions from the GAIA-DR2 \citep{GAIA2018} with 
radial velocities (RVs)  from \citet{maxted2008}, \citet{sacco2008} and \citet{hernandez2014}. For those stars with more than one measurements of RVs, we used a weighted mean to estimate the combined RV of the star.
We selected stars  with uncertainties in parallaxes below 20\%. 
Thus, distances  can be calculated as the inverse of the parallaxes. 
For stars with uncertainties larger than this value, we cannot apply the inverse relation between parallaxes and distances and the estimation of distances becomes an inference problem in which the use of prior assumptions is necessary \citep{bailer2015}. 
%Additionally, {from the radial velocity sample} we selected stars with distances between 300 pc and 500 pc. 
Additionally,
we selected stars with distances between 300 pc and 500 pc
{from the radial velocity sample}.
About 96\% (162/169) of the sample of known young stars with radial velocity measurements fulfill this criterion. 

Following \citet{jeffries2006}, we split the sample into two groups. One group (hereafter the \SOri population) has radial velocities between 27 km/s and 35 km/s and is
consistent with the radial velocity of the central system \citep[29.5 km/s;][]{kharchenko2007}. The other group (hereafter the sparse population) has radial velocities between 20 km/s and 27 km/s and could be older and closer than the first population  \citep{jeffries2006,maxted2008,hernandez2014}

Figure \ref{fig:RVdist} shows 
distances versus 
radial velocities (RVs) for the \SOri cluster
and the sparse populations in the
direction of the cluster. 
Using weighted means, 
distances and RVs were estimated for each population.
The \SOri po\-pu\-la\-tion has a 
distance of 401$^{+33}_{-28}$ pc and a 
RV of 31.1$\pm$1.3 km/s. The sparse population has a  distance of  382$^{+27}_{-23}$ pc and a 
RV of 23.6$\pm$1.4 km/s. { For SO411, there are several measurements of RVs: 22.2$\pm$1.1 km/s \citep{hernandez2014}, 21.7$\pm$1.5 km/s \citep{GAIA2018} and 22.9$\pm$0.5 km/s and 22.9$\pm$0.4 km/s \citep{kounkel2018}. Since, all these values are in agreement within the errors, it is unlikely that the smaller RVs of SO411 in comparison with the \SOri population are due to  the object being a spectroscopic binary. The weighted mean of RVs (22.8$\pm$1.2 km/s) is more than six $\sigma$ below the RVs expected for the \SOri population. Thus, it is ap\-par\-ent that the RVs of SO411 are in better agreement with the sparse population. On the other hand, the distance  of SO~411 (374$\pm$7 pc) is about one $\sigma$ below the distance estimated for the \SOri population and in better agreement with the sparse population. The difference in the distance distributions between the \SOri population and the sparse population is confirmed by a Kolmogorov-Smirnov (K-S) test, showing a significance level of only 0.3\%.

\begin{figure}[t!]
\plotone{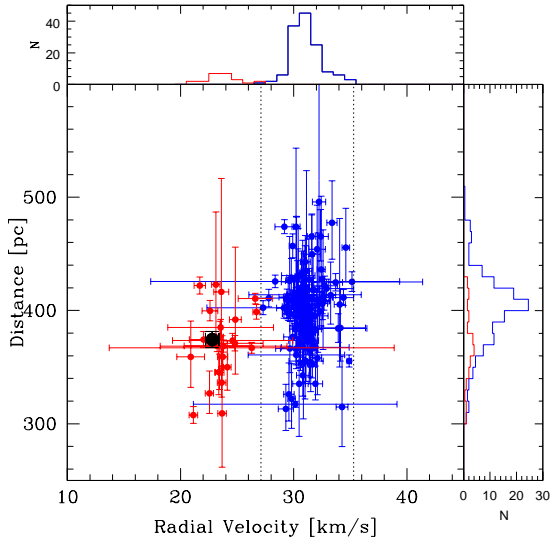}
\centering
\caption{Distance  vs. radial velocity for stars in the \SOri po\-pu\-la\-tion (blue dots) and the sparse population (red dots). The black dot indicates the location of the star SO~411. The dotted lines indicate the 3-$\sigma$ limits of the \SOri population. The upper panel and right panel show the RVs distribution and the distance distribution, respectively\label{fig:RVdist}} 
\end{figure}

Figure \ref{fig:PM} shows the vector point diagram and the proper motions distributions for the \SOri and the sparse po\-pu\-la\-ti\-ons.  Using weighted mean after applying 3-$\sigma$ clipping, we estimate the proper motions for the \SOri po\-pu\-la\-ti\-on to be $\mu_\alpha$= 1.5$\pm$0.5 and $\mu_\delta$= -0.6$\pm$0.5 and for the sparse population to be $\mu_\alpha$= 0.9$\pm$1.3 and $\mu_\delta$= -1.1$\pm$0.7. The proper motions of SO~411 ($\mu_\alpha$= 2.0$\pm$0.1 and $\mu_\delta$= -1.6$\pm$0.1) locate this object at $\sim$2.5$\sigma$ from the mean proper motions of the \SOri population.The difference between the \SOri population and the sparse population is confirmed by a K-S test, showing a significance level of 3\% and 0.008\% for the proper motion distributions of $\mu_\alpha$ and $\mu_\delta$, respectively.

Finally, \citet{caballero2008b} analyzed the spatial distribution of possible members of the \SOri cluster and suggests that this cluster has two components: a dense core that extends from the center to a radius of 20{\arcmin}, in which most members are located, and a rarefied halo at larger separations. SO~411 is located at 21.4{\arcmin} from the center, slightly beyond the dense core limit.

 In brief, the kinematic properties of SO~411 suggest that this star does not belong to the \SOri cluster and it is most likely associated with an older and sparser population. Distributions in distance also suggest that the sparse population is located in front of the cluster.
% {A larger scale study in this region also indicates the presence of two populations of young stars in the OB1b subassociation (which includes the SOri cluster), located at two different distances, which may be due to the OB1a subassociation overlapping on front of OB1b \citep{briceno2018}
 {A larger scale study  
in the Ori OB1 region, which includes the \SOri cluster
also indicates the presence of two populations of young stars, 
located at two different distances; specifically, the Ori OB1a subassociation is closer than Ori OB1b, but they spatially overlap on the plane of the sky  
\citep{briceno2018}}

\begin{figure}[t!]
\plotone{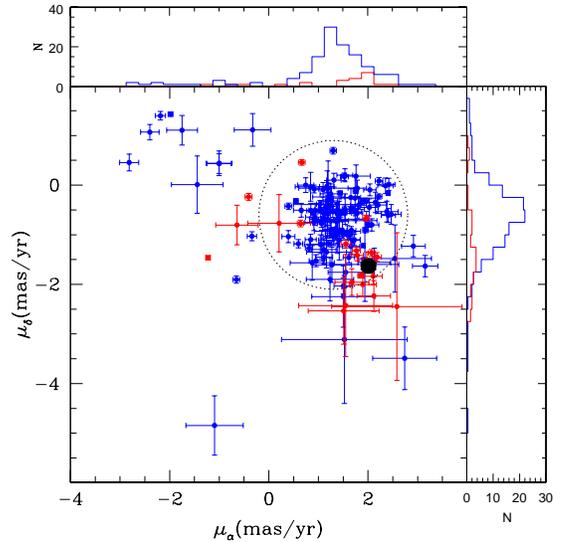}
\centering
\caption{Vector point diagram for the \SOri population (blue dots) and for the sparse population (red dots).  The black dot indicates the position of SO~411 which is close to the 2.5 $\sigma$ limit of the \SOri stellar population (dotted open circle).The upper panel and right panel show the proper motions distributions for right ascension and declination, respectively  \label{fig:PM}} 
\end{figure}

\subsection{Stellar properties and mass accretion rate} \label{sec:param}

The effective temperature (\teff), intrinsic colors and the bolometric correction of SO~411 were obtained by in\-ter\-po\-lat\-ing its spectral type in the table of \citet{pecaut2013} for pre-main sequence stars. Following the method described in \citet{hernandez2014} and using the colors {V-\Ic} and {V-J}, we calculated an extinction of 0.2 magnitudes in the visual band. Using this value and the reddening law from \citet[][$R_V$=3.1]{mathis1990}, the observational data described in \S\ref{sec:obs} were corrected for extinction.

Assuming a distance of 374$\pm$7 pc \citep{GAIA2018},
we found a stellar luminosity (L$_{*}$) of 9.1$\pm$0.5\lsun. The uncertainty in L$_{*}$ is calculated only from the visual photometric error. The stellar mass (M$_{*}$) and stellar radius (R$_{*}$) were estimated by interpolating the {\teff} and the {L$_{*}$} in the evolutionary tracks from \citet{siess2000}.

We estimated the accretion luminosity
(L$_{acc}$)
from several indicators. 
Table \ref{tab:star} shows the different values of $\dot{M}$ 
we obtained.
One
indicator is the U-band excess,
which measures the emission
shock at the stellar surface.
We used the correlation of this
excess with L$_{acc}$ from
\citet{gullbring1998},
and obtained the mass accretion
rate ($\dot{M}$) using
the stellar mass and radius.

The luminosity of the H$\alpha$
line 
(L$_{H\alpha}$)
can also be used as an accretion proxy \citep[e.g.][]{fang2009,ingleby2013,alcala2014,fairlamb2017}.
We estimated this luminosity from the H$\alpha$ equivalent width (\ewha), which is obtained by measuring the flux of the spectral feature normalized to the continuum level that is expected when interpolating between two adjacent continuum bands \citep{hernandez2004}. The intrinsic value for SO~411 (\ewha$_{std}$) is obtained by interpolating its spectral type in the values of {\ewha} obtained for the standard sequence in Figure \ref{fig:sptclass}.
Assuming no veiling at the con\-ti\-nuum of H$\alpha$, the corrected value  of SO~411 is obtained by \ewha$_{cor}$={\ewha}-\ewha$_{std}$. The flux of H$\alpha$ is calculated by 
F$_{H\alpha}$= \ewha$_{cor}$ $\times$ F$_{cont}$, where F$_{cont}$ is the continuum flux 
at 6563{\AA} obtained by interpolating between the fluxes measured in the filters {\Rc} and {\Ic}. The luminosity of H$\alpha$ follows from the assumed distance.
Several correlations between L$_{acc}$  and L$_{H\alpha}$ have been presented in the literature. The correlations by \citet{ingleby2013} and \citet{alcala2014}
were based on samples of CTTS, while the correlation by \citet{fairlamb2017} were 
based on a sample of HAeBe stars. 
On the other hand, the correlation presented by \citet{fang2009} is based on a sample that includes CTTS and few HAeBe stars.
The values of $\dot{M}$ obtained with the different correlations are shown
in Table \ref{tab:star}. The HAeBe correlation
results in a higher estimate of $\dot{M}$ 
than the CTTS correlations, a fact already noted
by \citet{fairlamb2017}. The luminosity of H$\alpha$
in SO~411 is $\sim 3.7 \times 10^{-3}$ \lsun,
which puts the object in between the 
lower L$_{H\alpha}$ values of the CTTS and
the higher values of the HAeBe, so it is not clear which of the two correlations between
L$_{acc}$  and L$_{H\alpha}$ is appropriate for SO~411.
 Given this uncertainty, we adopted as the
mass accretion rate of SO411 the weighted mean 
of the values obtained from the correlations
between L$_{acc}$  and L$_{H\alpha}$ in Table \ref{tab:star}
(log($\dot{M}$[\msun yr$^{-1}$])=-8.3+-0.3). 
This value is consistent with the mass accretion
rates found in other transitional disks
\citep{espaillat2014}.

\begin{deluxetable*}{ccc}%[b!]
\tablecaption{Stellar Properties  \label{tab:star}}
\tablecolumns{3}
%\tablenum{2}
\tablewidth{0pt}
\tablehead{
\colhead{Parameter} & \colhead{Value} & \colhead{Comments}
}
\startdata
Spectral Type & F6.0 $\pm$ 1.0 & \nodata \\
\av & 0.2 $\pm$ 0.1  & \nodata \\
T$_{*}$ (K) & 6250  &  using \citet{pecaut2013} \\
L$_{*}$ (\lsun) & 9.1$\pm$0.5 & \nodata \\
M$_{*}$ (\msun) & 1.6  & using \citet{siess2000}\\
R$_{*}$ (\rsun) & 2.0  & using \citet{siess2000}\\
{ log($\dot{M}$ [\msun yr$^{-1}$])}  & {-8.9} & from U-band excess \citep{gullbring1998} \tablenotemark{a}\\
{ log($\dot{M}$ [\msun yr$^{-1}$])}  & { -8.5$\pm$0.3} & {from H$\alpha$ and the relation of \citet{fang2009}}\tablenotemark{a} \\
{ log($\dot{M}$ [\msun yr$^{-1}$])}  & { -8.9$\pm$0.3} & from H$\alpha$ and the relation of \citet{alcala2014}\tablenotemark{a}\\
{ log($\dot{M}$ [\msun yr$^{-1}$])}  & { -8.8$\pm$0.9} & from H$\alpha$ and the relation of \citet{ingleby2013} \tablenotemark{a}\\ % using Ingleby 2013
{ log($\dot{M}$ [\msun yr$^{-1}$])}  & { -8.0$\pm$0.1} & from H$\alpha$ and the relation of \citet{fairlamb2017}\tablenotemark{a}\\
\tableline
\enddata
\tablenotetext{a}{The uncertainties only reflect the propagation of the fitting parameter errors in the correlation between L$_{acc}$  and L$_{H\alpha}$}
\end{deluxetable*}

\subsection{Spectral Energy Distribution}

We constructed the spectral energy distribution 
of SO~411, shown in Figure \ref{fig:sed}.
We used the data described in
\S \ref{sec:obs}, 
and magnitudes and fluxes collected from the literature: 2MASS photometry from \citet[J, H and K bands;][]{cutri2003}, 
IRAC (3.6, 4.5, 5.8 and 8.0\micron) and MIPS(24\micron) Spitzer's photometry from \citet{hernandez2007}, WISE photometry (3.4, 4.6, 12 and 22\micron) from \citet{cutri2013}, 
AKARI's fluxes at 9 and 18{\micron} from \citet{ishihara2010} and AKARI's flux at 90{\micron} from \citet{yamamura2010}. To cover the millimeter range, we included fluxes at 850{\micron} and 1300{\micron} obtained with the Submillimeter Common User Bolometer Array (SCUBA-2) camera on the 15-m James Clerk Maxwell Telescope \citep{williams2013} and the 1330{\micron} continuum flux with the Atacama Large Millimeter-sub-millimeter Array \citep[ALMA;][]{ansdell2017}.
 Given the spatial resolution of the collected data, all these measurements include the integrated emission from the star-disk system.
For comparison,
Figure \ref{fig:sed} also shows the 
stellar photosphere,
calculated
using the intrinsic colors 
for a F6 main sequence star from \citet{kenyon1995},
normalized to the dereddened J band flux.
Since \citet{pecaut2013} only include 3 optical magnitudes (B, V and \Ic) we used the intrinsic colors from \citet{kenyon1995} which include more optical magnitudes (U, B, V, {\Rc} and \Ic) to represent the stellar photosphere. The differences between these two standard values are less than 4\% in the optical bands.

The SED of SO~411 in Figure \ref{fig:sed} shows little excess over the photosphere in the near-IR bands, but large excess in the mid-IR and longer wavelengths. This
indicates that the disk of SO~411 can be classified as a transitional disk, in agreement with the classification of TD/PTD based on photometry \citep{hernandez2014}.
The outer optically thick disk
is therefore truncated at some distance from the star \citep{espaillat2011}.
There still is
some small flux excess over the photosphere 
in the near-IR, which could come from either
optically thin dust in the inner
cavity or from an optically thick inner
disk remaining close to the star. In the
last case, the disk would be a pre-transitional
disk \citep{espaillat2007a,espaillat2011}.

\begin{figure*}
\centering
\includegraphics[width=0.75\textwidth]{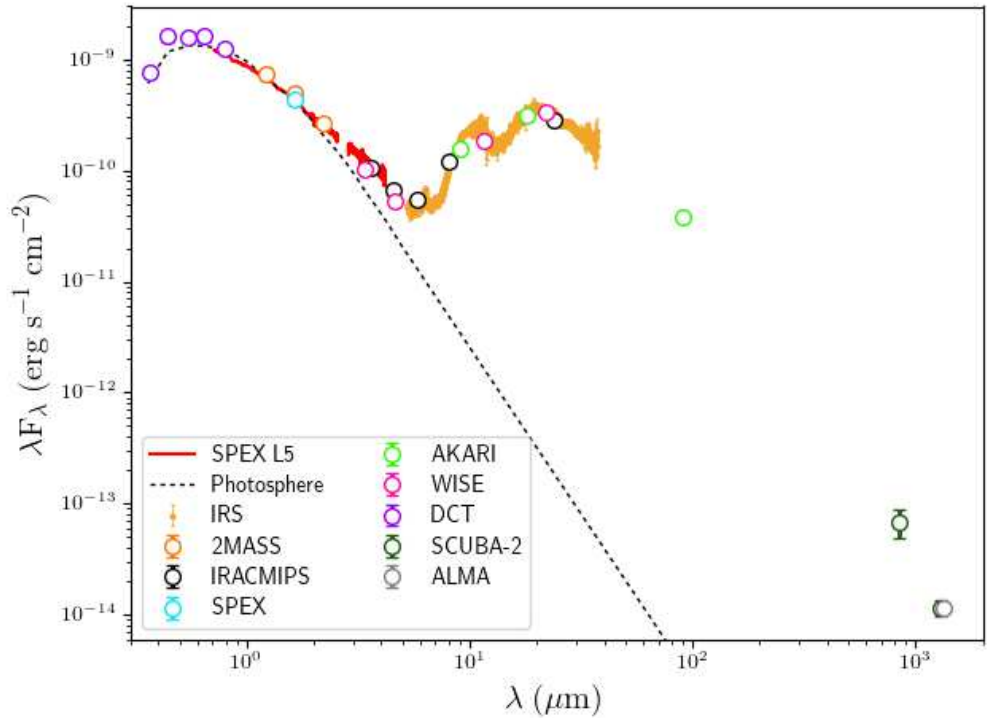}
\caption{ Spectral energy distribution of SO~411.
{ As reference, we show the 
stellar photosphere calculated
using the intrinsic colors 
for a F6 main sequence star from \citet{kenyon1995}, scaled at J.}
The sources for the
photometry and spectroscopy are
indicated in the legend.
\label{fig:sed}} 
\end{figure*}

\subsection{Veiling and NIR excess continuum} \label{sec:spex}

To elucidate the nature of the material
remaining inside the cavity
in the disk of SO~411,
we use the SpeX data to extract the
spectrum of the material emitting
in the near-IR.
Following  the methods in \citet{muzerolle2003} and \citet{espaillat2010}, 
we extracted 
the NIR excess continuum 
over the photosphere
in the SpeX spectrum.

We estimated the NIR excess continuum subtracting the spectroscopic template HD11443 
from the SPEX spectrum of SO~411. 
The spectrum of HD11443 (F6IV)
taken from
the IRTF Spectral Library\footnote{http://irtfweb.ifa.hawaii.edu/\~spex/IRTF\_Spectral\_Library/}\citep{cushing2005,rayner2009} was used as template.
Before the subtraction, the target 
was corrected by veiling  and both the target and the template spectra were
normalized to the K band.
A veiling of r$\sim$0.1 was estimated comparing the equivalent width of Br$\gamma$ measured for SO~411 with  
that of the template. 
Since Br$\gamma$ could have a non resolved emission contribution from the accretion shocks, the estimated veiling should be considered as an upper limit. Unfortunately, the SpeX spectrum was too noisy to estimate veiling from other spectroscopic lines not affected by accretion.

Figure \ref{fig:NIRexc} shows the NIR excess continuum for SO~411; 
for comparison, we also show the spectra of the pre-transitional
disk LkCa 15 and of the transitional disk
GM Aur \citep{espaillat2010}. 
Similarly to GM Aur, the NIR excess continuum of SO~411 is small and 
it is too broad to be fitted by a single blackbody, unlike
the spectrum of LkCa 15.
This e\-v\-idence suggests that the near-IR in SO~411
does not arise in optically thick emission near the
star but most likely in optically thin dust
inside the cavity, indicating that SO~411 is
not a pre-transitional disk.
Figure 5 also shows the presence of 
the PAH feature at 3.3{\micron}. Unfortunately this feature falls between two orders of the Spex spectrum (\S \ref{sec:pah}).

\begin{figure}
\plotone{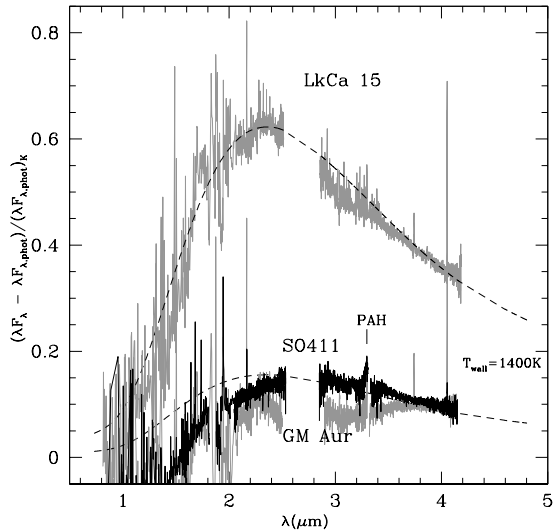}
\centering
\caption{NIR excess continuum. Here we compare the NIR excess continuum of SO~411 with that estimated for the stars LkCa15 and GM Aur \citep{espaillat2010}. SO~411 and GM Aur exhibit an IR excess continuum relatively small and flat which suggest that these stars are surrounded by a TD with optically thin dust inside the cavity. On the other hand, the PTD LkCa15 exhibits a relatively strong IR excess continuum that can be modeled as a single blackbody with temperature of 1400 K. The PAH feature at 3.3{\micron} falls between two SpeX orders (see \S\ref{sec:pah}). 
\label{fig:NIRexc}} 
\end{figure}

\subsection{Disk Model} \label{sec:model}

We calculated the structure and emission of the disk
surrounding SO~411 using the D'Alessio Irradiated Accretion Disks model \citep[DIAD;][]{dalessio1998,dalessio1999,dalessio2001, dalessio2005, dalessio2006}, following the methods described in \citet{espaillat2010},
that apply the DIAD model to stars surrounded by TD or PTD. Based on the analysis in \S \ref{sec:spex} we modeled the disk around SO~411 as a TD,
that is, a truncated optically thick disk, allowing
for 
optically thin dust inside the cavity. 
The input parameters that we held constant for the DIAD model were the accretion rate ($\dot{M}$) and the stellar properties  (Table \ref{tab:star}). Other parameters related to the disk structure
and to the 
dust mixtures in the different
parts of the disk
were
varied to achieve the best fit model to the observed SED. We assumed a value of 60{\degree} as the 
inclination of the system.

We assumed that the dust is thermally coupled to the gas and the dust size distributions
are proportional to $a^{-3.5}$, where $a$ is the radius of spherical grains, between $a_{min}$ (fixed at 0.005\micron) and $a_{max}$ \citep{mathis1977}.  
In the optically thick disk
the dust consists of silicates with mass fraction
relative to gas of $\zeta_{\rm sil}$=0.004 and 
graphite with mass fraction of $\zeta_{\rm acm}$=0.0025 \citep[$\zeta_{\rm std}$=0.0065;][]{mcclure2013}. 
The silicate grains
are composed of amorphous silicates ({pyroxene stoichiometry} and {olivine stoichiometry}) and crystalline silicates ({fosterite}, and {enstatite}). 
The opacities for the amorphous
silicates and graphite were computed using MIE theory with optical constants from \citet{dorschner1995} and \citet{draine1984}, respectively.
The opacities for 
the crystalline silicates were taken from  \citet{sargent2009b},
we used the Continuous Distribution 
of Ellipsoids (CDE) approximation to calculate the
opacities. We did not include water ice. 

The parameters of each disk component were adjusted simultaneously to fit the observed SED. The components of the disk model that contribute to the synthetic SED include 
the optically thick disk,
the inner edge of this disk (``the wall'',
located at $\rm{R_{w}}$),
and the optically thin region between the star and the wall. 
We ran a grid of more than 3000 models, varying the input parameters in the ranges
shown in Table \ref{tab:disk}, and calculated the $\chi^2$ comparing the synthetic SED with the observations. The theoretical
SED that best fitted the observations is shown 
in Figure \ref{fig:bestfit}.
The parameters of the model are shown in Table \ref{tab:mod_var}.
Additionally, we estimated confidence intervals for the abundance of the different silicates species, for the location and height of the disk inner edge and for the size of the outer disk. To set these intervals, we followed the procedure in \citet{mauco2016,mauco2018} to calculate the likelihood function, $\mathcal{L}$, of each parameter. The confidence intervals are given as those extreme limits at which the area below the likelihood
curve maximum is 63\% (1$\sigma$) of its total area \citep{sivia2012}. 
For those cases where the best value falls
on one of the edges of the range of values used in the models, we
have considered these as upper or lower limits, and they
are indicated by parenthesis instead of square brackets in
Table~\ref{tab:mod_var}. Figure \ref{fig:confidence} shows the intervals for the size of the outer disk as well as for the location and height of the disk inner edge (light-blue shaded regions). 
Details for each disk component are given below.

\begin{figure*}
\centering
\includegraphics[width=0.75\textwidth]{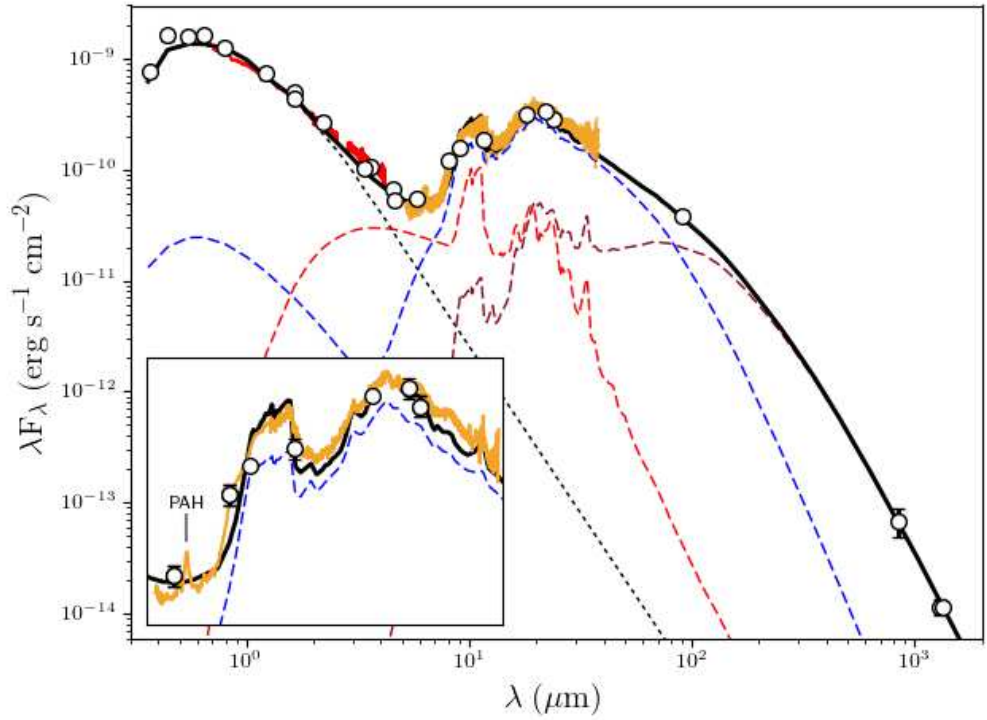}
\caption{
The observed SED of SO~411 and the best fit model. 
The synthetic SED (black solid line) includes the photosphere (black dotted line), the optically thin dust (red dashed line), the wall (blue dashed line) and the disk (brown dashed line). We also show in the figure inset the SED at the IRS wavelength range. 
\label{fig:bestfit}} 
\end{figure*}

\begin{figure*}
\centering
\includegraphics[width=0.9\textwidth]{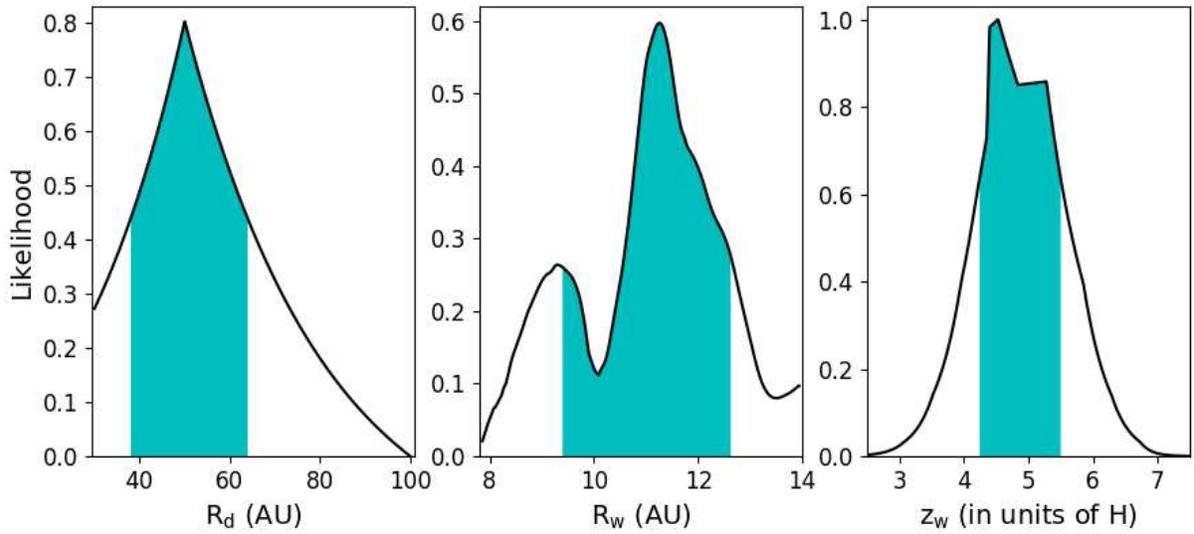}
\caption{
Likelihood function, $\mathcal{L}$, for the outer disk size (left), the location of the disk inner edge (middle) and its height (right). Confidence intervals are indicated by light-blue shaded regions in each panel and are defined as those enclosing 63\% (1$\sigma$) of the total area of $\mathcal{L}$. These intervals are reported in Table~\ref{tab:mod_var}.
\label{fig:confidence}} 
\end{figure*}

{\bf Outer disk edge:} We represent the inner edge
of the optically thick disk
by an optically thick ``wall'',
assumed to be vertical with evenly distributed dust. 
We calculated the structure and emission of this
wall following
\citet{dalessio2004}. 
The dust mixture, the  temperature of the wall, ${\rm T_w}$, the radius of wall, ${\rm R_w}$, 
and the height of the wall, ${\rm z_w}$, in units of the gas scale height ($H$) 
are varied to obtain the best model. 

The best fit parameters are given in Table \ref{tab:mod_var}, including the relative abundances of the silicate mixture and their uncertainties.
We found that the
emission from the wall atmosphere 
is the main
contributor to
the silicate bands observed in the IRS spectra
(cf. Figure \ref{fig:bestfit}).

{\bf Optically thick outer disk:} The emission of the 
optically thick disk was calculated using the prescription in \citet{dalessio2006}. The disk is heated by stellar irradiation and viscous dissipation and 
the disk structure and temperature distribution are calculated self-consistently
for each radii. To simulate grain growth and settling, 
we assume that the disk is composed of two grain size distributions. In the upper disk layers, the grains have small sizes (with $a_{max}$ < 5\micron) and in the disk midplane the maximum grain size is given by 
$a_{maxb}$, which we adopt as a parameter. 
The viscosity is 
written in terms of 
the  parameter $\alpha$
\citep{shakura1973,dalessio1998,dalessio1999}. 
Dust settling is parametrized with
the quantity
$\epsilon$=$\zeta_{\rm small}/\zeta_{\rm std}$, where $\zeta_{\rm small}$ is the dust-to-gas mass ratio of the small grain population
and $\zeta_{\rm std}$ represents the sum of the mass fraction of the different dust components relative to gas. 
{ We assume that the dust composition in the outer
disk is the same as that of its edge.}
%The dust composition was the same as in the outer disk edge.
The radius of the disk, ${\rm R_d}$, $\alpha$, $\epsilon$ and  $a_{maxb}$ are varied to achieve the best fit to the observed SED. 
The best fit parameters are given in Table \ref{tab:mod_var}.

The mass of the disk, $M_{\rm disk}$, is calculated by integrating
the surface density up to the disk radius;
it is also given in Table \ref{tab:mod_var}. Our $M_{\rm disk}$, which is calculated based on dust emission and assuming a dust-to-gas mass ratio of 0.0065, is greater (by a factor of $\sim$10) than the upper limit reported by \citet{ansdell2017} from CO observations. They stated, however, that their gas mass estimate has several caveats since it depends on the uncertain [CO]/[$\rm H_{2}$] and [CO]/[$\rm ^{13}CO$] ratios assumed in their models. If we, on the other hand, compare their dust mass estimate (from continuum observations at 1.33 mm) with ours, we found that our dust mass is only 3.5 greater than theirs. This difference is expected since they used the simplified approach of a single grain opacity in an isothermal disk, which tends to overestimate the grain opacity and hence, underestimates the mass. With our dust grain opacity, however, we can reproduce not only the flux at 1.33 mm but the entire SED.

{\bf Optically thin inner region:} 
The emission from 
dust 
inside the cavity was calculated by integrating the specific intensity
emerging from optically thin annuli,
in which all of the dust grains are heated by stellar radiation \citep{calvet2002,espaillat2010}. 
The maximum grain size ($a_{maxthin}$), the mass of the optically thin dust (${\rm M_{thindust}}$), the size of the region ($R_{\rm i,thin}$,$R_{\rm o,thin}$) and the fractional abundances of the silicates in the dust mixtures are varied to obtain the best model.
The optically thin dust is composed by $\sim$0.03 lunar masses distributed between 0.3 AU and 9.0 AU with a maximum grain size of 5\micron.
Following \citet{calvet2005,espaillat2007a,espaillat2007b} we add organic ($\zeta_{\rm org}$=0.001), troilite ($\zeta_{\rm troi}$=0.000768) and amorphous carbon ($\zeta_{\rm acm}$=0.001) to the dust mixture for comparison with other works in the literature. 
The opacities for organics, amorphous carbon and troilite are adopted from \citet{pollack1994} and \citet{begemann1994}. For the silicates we used $\zeta_{\rm sil}$=0.004. The parameters of the best fit model are shown in Table~\ref{tab:mod_var}.
We stress that this is a proxy of the composition and spatial distribution of the optically thin dust since we need high resolution, near infrared interferometry to trace this component in detail.  

\begin{deluxetable*}{cc}%[b!]
\tablecaption{Parameters explored  \label{tab:disk}}
\tablecolumns{2}
%\tablenum{2}
\tablewidth{0pt}
\tablehead{
\colhead{Parameter} & \colhead{Values} 
}
\startdata
\tableline
\colhead{Optically Thick Disk} \\
\tableline
$\epsilon$ & from 1 to 1e-5, step $\Delta$log($\epsilon$)=1  \\
$\alpha$  &  from 1 to 1e-4, step $\Delta$log($\alpha$)=1 \tablenotemark{a}\\
$a_{max}$($\mu m$) & from 0.1 to 1e4, step $\Delta$log($a_{maxw}$)=1 \tablenotemark{e} \\ %JH: this was fitted?
R$_{disk}$ (AU)  & from 50 to 300, step=50 \tablenotemark{b} \\
\tableline
\colhead{Optically Thick Wall} \\
\tableline
T$_{wall}$ (K)  & from 100 to 1400, step=100 \tablenotemark{c} \\
z$_{wall}$ (AU) & from 1.0 to 5.5, step=0.1 \\
$a_{maxw}$($\mu m$) & from 0.1 to 1e4, step $\Delta$log($a_{maxw}$)=1 \tablenotemark{e}\\
\tableline
\enddata
\tablenotetext{a}{We refine our best fit exploration including $\alpha$ of 0.003, 0.0006 and 0.0008}
\tablenotetext{b}{We refine our best fit exploration including steps of 10 AU.}
\tablenotetext{c}{We refine our best fit exploration including steps of 10 K.}
\tablenotetext{e}{We refine our best fit exploration including values of 0.25, 0.5, 0.75, 1.5, 2, 3, 4 and 5 \micron}
\end{deluxetable*}

\section{Discussion}

\subsection{Implications of dust mineralogy on disk evolution}\label{sec:sil}

Mid-infrared spectra of young stars can be well reproduced by a mixture of five main dust species: amorphous silicates with olivine and pyroxene stoichiometry, 
 and crystalline forsterite, enstatite and silica. Since the interstellar medium (ISM) is characterized by amorphous silicates, the crystalline grains found in HAeBe and TTS systems must have been produced in the disk.   Crystalline silicates such as forsterite and enstatite require formation temperatures above 1100 K \citep[e.g.,][]{fabian2000}. These temperatures are naturally reached near the central star, where one expects a large fraction of crystalline silicates. This has been confirm by interferometric observations of HAeBe disks showing highly crystalline inner disks \citep{vanBoekel2004}. However, Spitzer IRS spectra 
also indicate the presence of large amounts of warm $\sim$400-500 K crystalline silicates (enstatite or forsterite) at temperatures lower than the glass temperature of silicates
in the disks around a number
of young sources
\citep{sargent2009a}. 
This can be explained through transport of grains and radial mixing of material from hotter inner disk regions toward the cooler outer regions at larger radii from the central star \citep[e.g.,][]{gail2001}. However, \citet{bouwman2008} showed that the chemical gradients observed in circumstellar disks are difficult to explain in this scenario. 
Fast outward radial transport of crystals
during the outburst of EX Lup 
was found by
\citet{juhasz2012}, but it is not
clear if this mechanism would
reach beyond a few AU.
Therefore, silicate crystals in SO~411
%must 
most likely have formed locally in the disk, i.e. where they are presently observed.

In the case of SO~411, sharp bands of crystalline silicates can be observed in its IRS spectra.
 Modeling shows that
the inner 
edge of the truncated disk dominates the IRS emission (see Figure~\ref{fig:bestfit}), which indicates that thermal processing of dust has already taken place in
this region. 
The inset of Figure \ref{fig:bestfit} shows the IRS spectral range of SO~411 with the best fit model (solid line).  
As shown in Table \ref{tab:mod_var}, half the silicates of SO~411 are not amorphous, indicating a significant level of dust processing in its disk.

Enstatite and forsterite crystals have quite
distinct spectral features that allow
identification and quantification of their abundance. In particular, in the 10$\mu$m
region, enstatite has a conspicuous feature at 9.3 $\mu$m,
while forsterite strongest feature is at 10.8 $\mu$m
\citep[][Figure 1]{watson2009}.
The enstatite 9.3 $\mu$m feature in
the spectrum of SO~411 is very weak 
compared to forsterite (cf. Figure~\ref{fig:SO411_HB}),
indicating that forsterite is the most abundant silicate crystal (see Table~\ref{tab:disk}), with a forsterite-to-enstatite ratio of 4. Similarly, olivine (amorphous precursor of forsterite) is slightly more abundant than pyroxene (amorphous precursor of enstatite). Infrared observations of the upper layers of protoplanetary disks around HAeBe and TTS show a different trend, i.e., enstatite is seen to be more concentrated towards
the inner disk ($\lambda$ $\sim$ 10 $\mu$m, T $\sim$ 500 K), while forsterite is more abundant in the outer disk ($\lambda$ $\sim$ 30 $\mu$m, T $\sim$ 120 K) regions \citep{kessler2006, bouwman2008, meeus2009, juhasz2010}. According to \citet{juhasz2010}, under chemical equilibrium the crystal population is dominated by enstatite, assuming solar 
abundances.  Our finding contradicts this prediction, suggesting that the crystals we see in the spectra of SO~411 were formed under non-equilibrium conditions. 

In order to produce in situ formation of crystalline silicates at a few AU from the star, one needs to consider an alternative heating mechanism beside irradiation; since at these distances the stellar radiation field cannot heat the dust at high enough temperatures for crystallization to occur. Given the TD morphology of SO~411, models of local heating of dust and gas by shock waves driven by tidal interactions of a giant planet with the disk has been proposed \citep{desch2005,bouwman2010,juhasz2010,mulders2011}. A remarkable case is the HAeBe star HD 100546. This star also possesses a TD with forsterite grains in a disk wall located at $\sim$13 AU \citep{mulders2011}, and exhibits complex radial structures with appearance of spiral arms as well as detections of multiple companions in IR and sub(-mm) observations \citep[e.g.,][]{boccaletti2013,mulders2013,currie2014,currie2015,walsh2014,quanz2015,pinilla2015,wright2015,garufi2016,follette2017}. Another forsterite source is RECX 5, a M4 TTS also surrounded by a TD in the $\eta$ Cha cluster \citep{bouwman2010}. 

As stated in \citet{juhasz2010}, if the crystals are formed by shock heating in the outer disk, amorphous dust grains are then heated above the annealing temperature for a very short time, favoring crystallization \citep{harker_desch2002}. Over such a short timescale chemical equilibrium cannot be achieved and the resulting crystal product will be forsterite, independently of the starting stoichiometry of the amorphous particles.  This can explain the observed dominance of forsterite in the outer disk of SO~411. Therefore, we conclude, as in \citet{mulders2011}, that the high forsterite abundance in the disk wall of SO~411 is somewhat connected to the presence of a hole, and hence, to giant planet formation. Additionally, studies on annealing of amorphous grains found that amorphous enstatite is converted to crystalline forsterite at T $\le$ 1000 K \citep{fabian2000,roskosz2011}. These studies concluded that the mineralogy of silicate dust of solar composition should naturally be dominated by olivine for T $\le$ 1000 K, while above this temperature pyroxenes should dominate. Our result then suggests an upper limit for the annealing temperature of silicate crystals of 1000 K in the disk around SO~411.
 
The forsterite features of SO~411 are strikingly similar to those observed in the ISO spectrum of Comet Hale-Bopp \citep{crovisier2000} (see Figure~\ref{fig:SO411_HB}). {Since comets can retain in their composition signatures the chemical and physical conditions under which they formed \citep[e.g.,][]{willacy2015},} this provides a natural link to the composition of the solar nebula indicating that by $\sim$10 Myr \citep[the age of OB1a,][]{briceno2007}, the local dust composition of the proto-solar nebula must have already resembled that of SO~411.
It also raises the possibility that the disk around SO~411 contains a significant amount of comets, something already proposed for HD 100546 \citep{malfait1998}. This locates the formation of Hale-Bopp-like objects near the outer giant planets in protoplanetary disks, as in the case of HD 100546.

\begin{figure}
\centering
\includegraphics[width=0.5\textwidth]{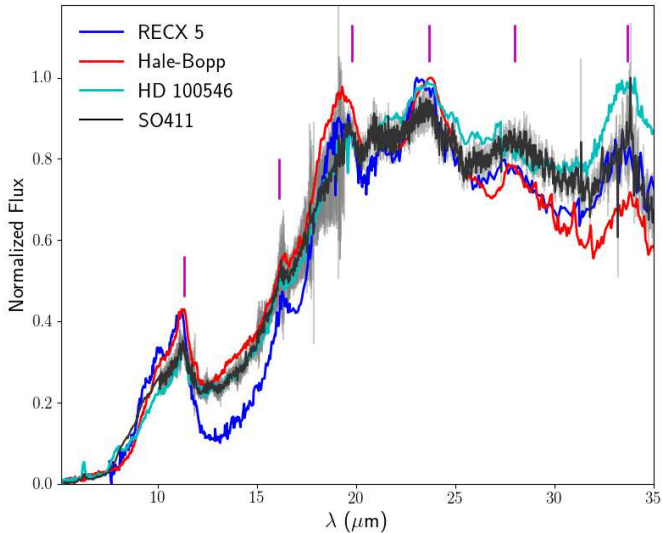}
\caption{Spitzer IRS spectra of SO~411 (black line) with errors (gray line). Also plotted are the ISO spectrum of comet Hale-Bopp (red line) and HD 100546 (light-blue line) as well as the IRS spectra of RECX 5 (blue line) for comparison. All the spectra have been normalized. The main forsterite features are also indicated (magenta vertical lines).   
\label{fig:SO411_HB}} 
\end{figure}

In contrast with HD 100546 and RECX 5 (also shown in Figure~\ref{fig:SO411_HB}, note 
%resembling of SO~411 with HD 100546), 
the similarity of SO~411 with HD 100546),
SO~411 also exhibits significant amounts of silica in its dust mixture. Laboratory annealing experiments of amorphous silicates also show that during the formation of forsterite, silica can be produced \citep[e.g.,][]{fabian2000}. In fact, silica has been found in the spectrum of young stars both in amorphous and in crystalline form \citep[e.g.,][]{bouwman2001,vanBoekel2005,sargent2009a}. As noted in \citet{bouwman2008}, the crystalization of forsterite from an amorphous enstatite precursor leads to a silica by-product. Therefore, the amount of silica should then be commensurate with the amount of forsterite produced from the enstatite component. This is exactly what we observe in SO~411. The reason why SO~411 does show silica content while others forsterite sources, such as HD 100546 and RECX 5 (and similarly, comet Hale-Bopp) do not, is unknown and a matter for future discussion.

\begin{deluxetable}{lrr}

\tablecaption{Disk Parameters \label{tab:mod_var}}
\tablecolumns{3}

\tablehead{
\colhead{Parameter} & \colhead{Best} & \colhead{Confidence}\\
\colhead{ } & \colhead{value} & \colhead{interval}
}
\startdata
\multicolumn{3}{c}{Optically Thick Outer Disk}\\ \noalign{\smallskip}
\hline
$\dot{M}$ (\msun yr$^{-1}$)........& 5 $\times 10^{-9}$ & --\\ 
$\alpha$ ..........................& 0.003 & --\\ 
$\epsilon$ ........................& 0.1 & --\\
$R_{\rm d}$ (au) ..................& 50 & [38 - 64]\\
$a_{\rm max}$ ($\mu$m) ............& 0.25 & -- \\
$a_{\rm maxb}$ ($\mu$m) ...........& 1000 & --\\
$\rm oliv_{ab}$ (\%)...............& 30 & [23 - 30)\\
$\rm py_{ab}$ (\%).................& 20 & [9.5 - 29]\\
$\rm forst_{ab}$ (\%)..............& 20 & [18 - 20)\\
$\rm enst_{ab}$ (\%)...............& 5  & (5 - 5.5]\\
$\rm silica_{ab}$ (\%) ............& 25 & (25 -35]\\
$M_{\rm disk}$ ($M_{\odot}$).......& 0.039 & --\\
\hline
\multicolumn{3}{c}{Optically Thick Outer Wall (*)}\\ \noalign{\smallskip}
\hline
$\rm amax_{w}$ ($\mu$m) ...........& 1.0 & -- \\
$T_{\rm w}$ (K) ...................& 260 & --\\
$R_{\rm w}$ (au) ..................& 11 & [9 - 13]\\
$z_{\rm w}$ (in units of $H$) .....& 4.4 & [4.2,5.5] \\
%$\rm oliv_{ab}$ (\%)...............& 30 & [23 - 30)\\
%$\rm py_{ab}$ (\%).................& 20 & [9.5 - 29]\\
%$\rm forst_{ab}$ (\%)..............& 20 & [18 - 20) \\
%$\rm enst_{ab}$ (\%)...............& 5 & (5 - 5.5]\\
%$\rm silica_{ab}$ (\%) ............& 25 & (25 -35]\\
\hline
\multicolumn{3}{c}{Optically Thin Region}\\ \noalign{\smallskip}
\hline
$R_{\rm i,thin}$ (au)  ............& 0.3 & --\\
$R_{\rm o,thin}$ (au) .............&9.0 & --\\
$\rm amax_{thin}$ ($\mu$m) ........& 5.0 & --\\
$\rm oliv_{ab}$ (\%) ..............& 99 & [94.6 - 99.6]\\
$\rm enst_{ab}$(\%)................& 0.2 & [0.02 - 2]\\
$\rm forst_{ab}$(\%)...............& 0.8 & [0.34 - 3.4]\\
$M_{\rm thindust}$ (lunar masses)..& 0.03 & --\\
\enddata
* { The optically thick outer disk and the optically thick outer wall have the same dust composition}
\tablenotetext{~}{- The confident intervals represent the 63\% ($\sim$ 1 $\sigma$) of the area below the likelihood curve maximum.} 
\tablenotetext{~}{ - The parenthesis represent those cases where the best value falls on one of the edges of the range of values used in the models.}
\end{deluxetable}

\subsection{PAH features} \label{sec:pah}

The detection of PAH emission is an indicator that carbonaceous material subsists in protoplanetary disks \citep{henning2011}. More than half of the HAeBe stars exhibit PAH features at 3.3, 6.2, 7.7, 8.6, 11.3 and 12.7 {\micron} \citep{acke2010, henning2011, seok2017}. On the other hand, PAH detections around T Tauri stars are more scarce;  
as far as we know, there are only three stars with spectral type K5 or later with reliable detection of PAH features: T54 \citep[K5;][]{espaillat2017}, EC 82 (K7) and IC348 LRL110 \citep[M0;][]{seok2017}\footnote{the stars IC348 LRL190(~M3) and J1829070+003838(K7) listed in \citep{seok2017} have marginal detections}. 
This suggests that a stellar UV radiation field is necessary to excite the PAH molecules \citep{geers2006,acke2010}. 
However, the few PAH detections in T Tauri stars suggest that visible and NIR radiation 
may be capable of exciting PAH molecules \citep{li2002,mattioda2005}. 
If UV radiation is not necessary to excite PAH molecules, the lack of PAH detection could be explained 
if the continuum and the 10{\micron} silicate emission feature hide the PAH features \citep{espaillat2017}.

PAH features are seen in the
IRS spectrum of SO~411 (Figure \ref{fig:bestfit}).
The 3.3{\micron} PAH feature is 
also hinted 
in the SpeX spectra (Figure \ref{fig:NIRexc}), but unfortunately part of the feature fell between two SpeX orders so we will
not discuss it further.
Although a
feature at 6.2 {\micron} is clearly
seen in the IRS spectrum,
the features at longer wavelength need to be extracted
from the
strong silicate features 
(Figure \ref{fig:bestfit}). The strong contribution
of crystalline silicates makes this task even more
difficult, specially at wavelengths longer than 10 {\micron},
so we confine our analysis to the features at
6.2 and 7.7 {\micron},
due to the CC stretch modes, and the
8.6 {\micron} feature arising in the CH bending mode.
We extracted the features following the procedure
of \citet{seok2017}; namely, we fitted the
``continuum'' with a cubic spline going through
the same anchor points as in that paper,
and then subtracted this continuum from the observed
spectrum. The resultant ``normalized'' spectrum
is shown in Figure \ref{fig:pah}. We decided to 
follow this procedure, instead of subtracting the
model spectrum, because we wanted to compare 
properties of the extracted PAH spectra of SO~411
to those of the extended set of Herbig Ae/Be and
T Tauri stars
in 
\citet{sloan2005}, \citet{acke2010} and \citet{seok2017}, extracted
with the same procedure.

In the selected wavelength range, besides the 6.2 {\micron} feature,
we detect a wide
feature around $\sim$ 8.3 {\micron},
which we identify with the 7.7 and 8.6 {\micron} features (Figure \ref{fig:pah}).
Red-shifted 6.2 and 7.7{\micron} PAH features have been observed in young stellar objects 
and the shifts
give clues about the
properties of the PAH emitters \citep[e.g.,][]{sloan2005,acke2010,seok2017}. 
To estimate the location of the peaks, we fitted the
features with Gaussian profiles, as shown in Figure
\ref{fig:pah}. We found that the 
6.2 {\micron} feature is
shifted to $\lambda_{6.2}$ = 6.27 {\micron}
and the 7.7 {\micron} feature to
$\lambda_{7.7}$ = 8.14 {\micron}.
These shifts are
consistent with the trend observed
for the set of 53 Herbig Ae
stars in \citet{acke2010}, who
find that 
the 6.2 {\micron} feature
can be shifted to a maximum value
of $\sim$ 6.275 {\micron}; for this
maximum
value, the corresponding shift in the
7.7 {\micron} feature ranges
between 7.82 to 8.1 {\micron}, in
agreement with the shifts
observed in SO~411.
The shift of the 7.7 {\micron} peak 
is consistent with the trend of
position versus effective temperature
found by \citet{acke2010}
for a sample of Herbig Ae and T Tauri
stars;
they find that the position of the peak
is $<$ 8.1 {\micron}
for Herbig Ae, while it
is $>$ 8.21 {\micron} for
TTS, so the observed
shift of 8.14 {\micron} in SO~411 
is completely consistent with
its intermediate nature between
these two sets of stars.

\citet{seok2017} explain the decrease
of $\lambda_{7.7}$ between
TTS and Herbig Ae
as the result of a decrease of
the size of the PAHs 
as the effective temperature
and
luminosity of the star 
increase. In turn,
the loss of small
grains in the Herbig Ae
could be explained
in terms of a higher
photodissociation rate for
small PAH when the number of
energetic UV photons increases
as the star gets hotter and
more luminous \citep{seok2017}.
To estimate the PAH size in SO~411
we follow
\citet{seok2017}, who characterize
the grain size by the parameter
a$_p$, which is the size that
corresponds to the maximum of the
log size distribution weighted
by mass. Using the fits to the
observed correlations of $<$a$_p>$ with
effective temperature
and luminosity to mass ratio,
we get 
$<$a$_p>$ = 5.1 and 4.6 {\AA}
for the values of T$_{eff}$ and L$_{*}$/M$_{*}$
of SO~411 (Table \ref{tab:star}), respectively.
Moreover, the value of
$<$a$_p>$ predicted from the
location of the 7.7 {\micron}
is 4.3 {\AA} \citep{seok2017}. Given the large
scatter in the calibrations and the
uncertainties in the location
of the continuum, we consider that
these values are consistent
with each other. A value of
$<$a$_p>$ between 4.5 and 5 {\AA}
is smaller than in the ISM but in the
upper end of the TTS distribution in the sample
analyzed in \citet{seok2017},
and indicates that small PAHs
still remain in the disk of SO~411.
Therefore, the
PAH spectrum of
SO~411 is consistent with
its intermediate state between
the hot and
luminous Herbig Ae and the less
massive and cooler T Tauri stars.

\citet{acke2010} find that
the mass fraction of small
silica grains in cold dust
decreases as the FWHM of the
6.2 {\micron} feature increases in their
sample of Herbig Ae. Our
measurement for the FWHM 
of this feature is 0.182 {\AA},
which would imply a negligible
silica mass fraction, according
to this correlation. In contrast, we
find a silica fraction of 25\%
of all silicates (\S \ref{sec:sil}).
Herbig Ae stars do not seem to have
abundances of silica
as high as that found in SO~411,
but they do occur in 
T Tauri stars \citep{sargent2009b}.
As \citet{acke2010} state,
the correlation between the
FWHM of the 6.2 {\micron} and the
silica abundance is unexpected. 
Our finding that the correlation
does not hold for less energetic
environments may give an additional
clue to its origin.

\begin{figure}[t]
\centering
\includegraphics[width=0.50\textwidth]{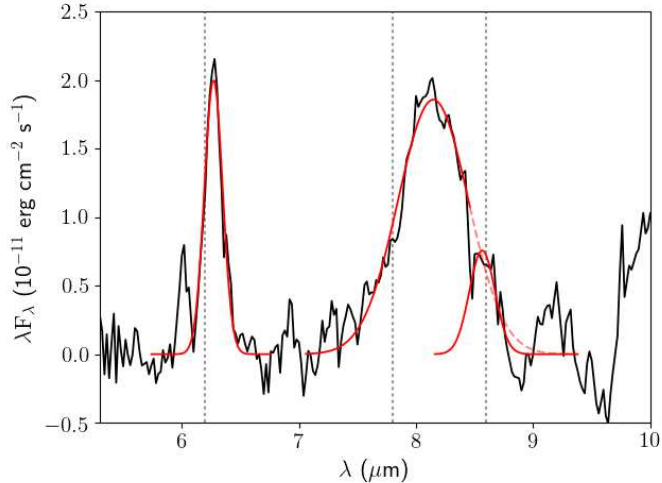}
\caption{Continuum-subtracted PAH spectrum for SO~411 (black solid line). Dotted vertical lines indicate the PAH features at 6.2$\mu m$,
7.7$\mu m$ 
and 8.6$\mu m$. Red lines indicate the Gaussian fit we used for each PAH features.
\label{fig:pah}}
\end{figure}

\section{Summary and conclusions} \label{sec:summ}

We present a detailed irradiated disk model for the intermediate mass T Tauri star SO~411 (1.6 \msun, spectral type F6). Our findings are as follows:

\begin{enumerate}

\item
Combining data from the GAIA-DR2 and available radial velocities, we find that SO~411 
is not associated with the \SOri cluster, but 
it is more likely related to an older and sparser population located in front of the cluster.

\item
The star is surrounded by a primordial disk with mass 0.039 \msun, with a gap extending to 11 AU from the star.
Except at the midplane and at the wall, the disk is populated by ISM grains
with a depletion of only 10\% relative to the standard dust-to-gas mass ratio.

\item
We find evidence that the star is still accreting, with a mass accretion rate of $ \dot{M} \sim 5 \times 10^{-9}$ \msun yr$^{-1}$.
Given the disk mass and mass accretion
rate, the most likely cause for the inner disk clearing is the interaction with a forming planet
\citep[cf.][]{espaillat2014}. 

\item
The SED shows a small near-IR excess
arising from optically thin emission
over a range of temperatures.
With this, 
we show that 
SO~411 is 
not surrounded by a pre-transitional disk.
The optically thin emission comes
from a small amount of micron size grains of 
0.03 lunar masses
located between 0.3 and 9 AU from the star. 

\item
Analysis of the 10$\mu$m silicate
feature profile indicates an 
approximately equal amount of
crystalline and amorphous material, with forsterite as the most abundant silicate crystal. 
The crystalline material is found on the wall of the outer disk,
which is the main contributor to the silicate features.
Shock waves induced by giant planets can locally heat the material in the wall of the disk to produce crystals. The forsterite abundance found in SO~411 points to crystal formation in non-equilibrium conditions.

\item
PAH features are detected in the IRS spectrum of SO~411. Analysis of the
6.2 and 7.7 {\micron} features
indicates that small PAHs remain
in the disk of SO~411.
The
PAH spectrum of
SO~411 is consistent with
its intermediate state between
the hot and
luminous Herbig Ae and the less
massive and cooler T Tauri stars.

\end{enumerate}

\vskip 0.2in

{\bf Acknowledgments.}
We are grateful to Jacques Crovisier for sending 
spectra of comet Hale-Bopp.
The authors acknowledge support from programs
UNAM-DGAPA-PAPIIT IN103017 and IN110816, Mexico.
This paper utilizes the \citet{dalessio1998,dalessio1999, dalessio2001,dalessio2005,dalessio2006} Irradiated Accretion Disk code. We wish to recognize the work of Paola D'Alessio, who passed away in 2013. Her legacy and pioneering works live on through her substantial contributions to the field.
{ An anonymous referee provided many
insightful comments that improved
the content and presentation of this work.}
We thank the institutions and
personnel that support data acquisition at the Observatorio Astron\'{o}mico Nacional at San Pedro Martir, the Lowell Observatory and the Mauna Kea Observatory.
This work is based on observations made with the Spitzer Space Telescope, which is operated by the Jet Propulsion Laboratory, under a contract with NASA.  We used the Infrared Telescope Facility, which is operated by the University of Hawaii under contract NNH14CK55B with the NASA.
These results made use of the Discovery Channel Telescope at Lowell Observatory, 
supported by Discovery Communications, Inc., Boston University, the University of Maryland,
the University of Toledo and Northern Arizona University. This publication makes use of data products from the Two Micron All Sky Survey, which is a joint project of the University of Massachusetts and the Infrared Processing and Analysis Center/California Institute of Technology, funded by the National Aeronautics and Space Administration and the National Science Foundation. We also make use of data products from the Wide-field Infrared Survey Explorer, which is a joint project of the University of California, Los Angeles, and the Jet Propulsion Laboratory/California Institute of Technology, funded by the National Aeronautics and Space Administration. This work is based on observations with AKARI, a JAXA project with the participation of ESA. %J.B.-P. acknowledges UNAM-PAPIIT grant no. IN110816 and UNAM's DGAPA. 
The grid of accretion disk models used in the present work has been calculated using the supercomputer Mouruka, at IRyA, provided by CONACyT grant number INFR-2015-01-252629.

\software{Astropy \citep{astropy2013}, Matplotlib \citep{matplotlib2005}, IRAF \citep{iraf1999}, SMART \citep{smart2004}, SpexTool \citep{cushing2004}}

%\textcolor{red}{YO CREO QUE FACILITIESSE REFIERE A AQUELLAS QUE FUERON USADASPARA OBTENER LAS NUEVAS OBSERVACIONES PRESENTADAS EN ESTE ARTICULO, NO HA TODAS LAS QUE HAN  OBTENIDO OBSERVACIONES DEL OBJETO. ES DECIR SERIA DCT Y IRTF (SpeX)}
\facilities{DCT(LMI), Spitzer(IRS), IRTF, SPM(2m-B\&C)}


\begin{thebibliography}{}

\bibitem[Acke \& van den Ancker(2004)]{AckeAncker04} Acke, B., \& van den Ancker, M.~E.\ 2004, \aap, 426, 151 

\bibitem[Acke et al.(2010)]{acke2010} Acke, B., Bouwman, J., Juh{\'a}sz, A., et al.\ 2010, \apj, 718, 558 

\bibitem[Alcal{\'a} et al.(2014)]{alcala2014} Alcal{\'a}, J.~M., Natta, A., Manara, C.~F., et al.\ 2014, \aap, 561, A2

\bibitem[Alexander et al.(2014)]{alexander2014} Alexander, R., Pascucci, I., Andrews, S., Armitage, P., \& Cieza, L.\ 2014, Protostars and Planets VI, 475

\bibitem[ALMA Partnership et al.(2015)]{alma2015} ALMA Partnership, Brogan, C.~L., P{\'e}rez, L.~M., et al.\ 2015, \apjl, 808, L3 

\bibitem[Andrews et al.(2016)]{andrews2016} Andrews, S.~M., Wilner, D.~J., Zhu, Z., et al.\ 2016, \apjl, 820, L40

\bibitem[Ansdell et al.(2017)]{ansdell2017} Ansdell, M., Williams, J.~P., Manara, C.~F., et al.\ 2017, \aj, 153, 240

\bibitem[Andrews et al.(2011)]{andrews2011} Andrews, S.~M., Wilner, D.~J., Espaillat, C., et al.\ 2011, \apj, 732, 42 

\bibitem[Astropy Collaboration et al.(2013)]{astropy2013} Astropy Collaboration, Robitaille, T.~P., Tollerud, E.~J., et al.\ 2013, \aap, 558, A33 

\bibitem[Bailer-Jones(2015)]{bailer2015} Bailer-Jones, C.~A.~L.\ 2015, \pasp, 127, 994 

\bibitem[Begemann et al.(1994)]{begemann1994} Begemann, B., Dorschner, J., Henning, T., Mutschke, H., \& Thamm, E.\ 1994, \apjl, 423, L71

\bibitem[Birnstiel et al.(2012)]{birnstiel2012} Birnstiel, T., Andrews, S.~M., \& Ercolano, B.\ 2012, \aap, 544, A79 

\bibitem[Boccaletti et al.(2013)]{boccaletti2013} Boccaletti, A., Pantin, E., Lagrange, A.-M., et al.\ 2013, \aap, 560, A20 

\bibitem[Bouwman et al.(2010)]{bouwman2010} Bouwman, J., Lawson, W.~A., Juh{\'a}sz, A., et al.\ 2010, \apjl, 723, L243 

\bibitem[Bouwman et al.(2008)]{bouwman2008} Bouwman, J., Henning, T., Hillenbrand, L.~A., et al.\ 2008, \apj, 683, 479-498 

\bibitem[Bouwman et al.(2001)]{bouwman2001} Bouwman, J., Meeus, G., de Koter, A., et al.\ 2001, \aap, 375, 950 

\bibitem[Brown et al.(2007)]{brown2007} Brown, J.~M., Blake, G.~A., Dullemond, C.~P., et al.\ 2007, \apjl, 664, L107

\bibitem[Brice{\~n}o et al.(2007)]{briceno2007} Brice{\~n}o, C., Hartmann, L., Hern{\'a}ndez, J., et al.\ 2007, \apj, 661, 1119 

\bibitem[Brice{\~n}o et al.(2018)]{briceno2018} Briceno, C., Calvet, N., Hernandez, J., et al.\ 2018, arXiv:1805.01008

\bibitem[Caballero(2008)]{caballero2008} Caballero, J.~A.\ 2008, \aap, 478, 667

\bibitem[Caballero(2008)]{caballero2008b} Caballero, J.~A.\ 2008b, \mnras, 383, 375 

\bibitem[Calvet et al.(2002)]{calvet2002} Calvet, N., D'Alessio, P., Hartmann, L., et al.\ 2002, \apj, 568, 1008 

\bibitem[Calvet et al.(2004)]{calvet2004} Calvet, N., Muzerolle, J., Brice{\~n}o, C., et al.\ 2004, \aj, 128, 1294

\bibitem[Calvet et al.(2005)]{calvet2005} Calvet, N., D'Alessio, P., Watson, D.~M., et al.\ 2005, \apjl, 630, L185 

\bibitem[Calvet \& D'Alessio(2011)]{calvet2011} Calvet, N., \& D'Alessio, P.\ 2011, Physical Processes in Circumstellar Disks around Young Stars, 14

\bibitem[Casassus(2016)]{casassus2016} Casassus, S.\ 2016, \pasa, 33, e013 

\bibitem[Chiang \& Murray-Clay(2007)]{chiang2007} Chiang, E., \& Murray-Clay, R.\ 2007, Nature Physics, 3, 604 

\bibitem[Cieza et al.(2013)]{cieza2013} Cieza, L.~A., Olofsson, J., Harvey, P.~M., et al.\ 2013, \apj, 762, 100

\bibitem[Crovisier et al.(2000)]{crovisier2000} Crovisier, J., Brooke, T.~Y., Leech, K., et al.\ 2000, Thermal Emission Spectroscopy and Analysis of Dust, Disks, and Regoliths, 196, 109 

\bibitem[Currie et al.(2015)]{currie2015} Currie, T., Cloutier, R., Brittain, S., et al.\ 2015, \apjl, 814, L27 

\bibitem[Currie et al.(2014)]{currie2014} Currie, T., Muto, T., Kudo, T., et al.\ 2014, \apjl, 796, L30 

\bibitem[Cushing et al.(2004)]{cushing2004} Cushing, M.~C., Vacca, W.~D., \& Rayner, J.~T.\ 2004, \pasp, 116, 362 

\bibitem[Cushing et al.(2005)]{cushing2005} Cushing, M.~C., Rayner, J.~T., \& Vacca, W.~D.\ 2005, \apj, 623, 1115

\bibitem[Cutri et al.(2003)]{cutri2003} Cutri, R.~M., Skrutskie, M.~F., van Dyk, S., et al.\ 2003, VizieR Online Data Catalog, 2246

\bibitem[Cutri et al.(2013)]{cutri2013} Cutri, R.~M., et al.\ 2013, VizieR Online Data Catalog, 2328,

\bibitem[Currie et al.(2009)]{currie2009} Currie, T., Lada, C.~J., Plavchan, P., et al.\ 2009, \apj, 698, 1 

\bibitem[D'Alessio et al.(1998)]{dalessio1998} D'Alessio, P., Cant{\"o}, J., Calvet, N., \& Lizano, S.\ 1998, \apj, 500, 411 

\bibitem[D'Alessio et al.(1999)]{dalessio1999} D'Alessio, P., Calvet, N., Hartmann, L., Lizano, S., \& Cant{\'o}, J.\ 1999, \apj, 527, 893 

\bibitem[D'Alessio et al.(2001)]{dalessio2001} D'Alessio, P., Calvet, N., \& Hartmann, L.\ 2001, \apj, 553, 321 

\bibitem[D'Alessio et al.(2004)]{dalessio2004} D'Alessio, P., Calvet, N., Hartmann, L., Muzerolle, J., \& Sitko, M.\ 2004, Star Formation at High Angular Resolution, 221, 403 

\bibitem[D'Alessio et al.(2005)]{dalessio2005} D'Alessio, P., Hartmann, L., Calvet, N., et al.\ 2005, \apj, 621, 461 

\bibitem[D'Alessio et al.(2006)]{dalessio2006} D'Alessio, P., Calvet, N., Hartmann, L., Franco-Hern{\'a}ndez, R., \& Serv{\'{\i}}n, H.\ 2006, \apj, 638, 314 

\bibitem[D'Alessio(2009)]{dalessio2009} D'Alessio, P.\ 2009, Revista Mexicana de Astronomia y Astrofisica Conference Series, 35, 33 

%\bibitem[Dullemond \& Dominik(2004)]{dullemond2004} Dullemond, C.~P., \& Dominik, C.\ 2004, \aap, 417, 159 

\bibitem[Desch et al.(2005)]{desch2005} Desch, S.~J., Ciesla, F.~J., Hood, L.~L., \& Nakamoto, T.\ 2005, Chondrites and the Protoplanetary Disk, 341, 849 

\bibitem[Dong \& Fung(2017)]{dong2017} Dong, R., \& Fung, J.\ 2017, \apj, 835, 146

\bibitem[Dorschner et al.(1995)]{dorschner1995} Dorschner, J., Begemann, B., Henning, T., Jaeger, C., \& Mutschke, H.\ 1995, \aap, 300, 503 

\bibitem[Draine \& Lee(1984)]{draine1984} Draine, B.~T., \& Lee, H.~M.\ 1984, \apj, 285, 89 

\bibitem[Dullemond \& Dominik(2004)]{dullemond2004} Dullemond, C.~P., \& Dominik, C.\ 2004, \aap, 421, 1075 

\bibitem[Dullemond \& Dominik(2005)]{dullemond2005} Dullemond, C.~P., \& Dominik, C.\ 2005, \aap, 434, 971 

\bibitem[Espaillat et al.(2007a)]{espaillat2007a} Espaillat, C., Calvet, N., D'Alessio, P., et al.\ 2007, \apjl, 664, L111

\bibitem[Espaillat et al.(2007b)]{espaillat2007b} Espaillat, C., Calvet, N., D'Alessio, P., et al.\ 2007, \apjl, 670, L135 


\bibitem[Espaillat et al.(2010)]{espaillat2010} Espaillat, C., D'Alessio, P., Hern{\'a}ndez, J., et al.\ 2010, \apj, 717, 441 

\bibitem[Espaillat et al.(2014)]{espaillat2014} Espaillat, C., Muzerolle, J., Najita, J., et al.\ 2014, Protostars and Planets VI, 497 

\bibitem[Espaillat et al.(2017)]{espaillat2017} Espaillat, C.~C., Ribas, {\'A}., McClure, M.~K., et al.\ 2017, \apj, 844, 60

\bibitem[Espaillat et al.(2008)]{CE_08b} Espaillat, C., Muzerolle, J., Hern{\'a}ndez, J., et al.\ 2008, \apjl, 689, L145 

\bibitem[Espaillat et al.(2011)]{espaillat2011} Espaillat, C., Furlan, E., D'Alessio, P., et al.\ 2011, \apj, 728, 49 

\bibitem[Espaillat et al.(2012)]{CE_12} Espaillat, C., Ingleby, L., Hern{\'a}ndez, J., et al.\ 2012, \apj, 747, 103 

\bibitem[Ercolano \& Pascucci(2017)]{ercolano2017} Ercolano, B., \& Pascucci, I.\ 2017, Royal Society Open Science, 4, 170114

\bibitem[Fabian et al.(2000)]{fabian2000} Fabian, D., J{\"a}ger, C., Henning, T., Dorschner, J., \& Mutschke, H.\ 2000, \aap, 364, 282 

\bibitem[Fabricant et al.(1998)]{fabricant1998} Fabricant, D., Cheimets, P., Caldwell, N., \& Geary, J.\ 1998, \pasp, 110, 79 

\bibitem[Fairlamb et al.(2017)]{fairlamb2017} Fairlamb, J.~R., Oudmaijer, R.~D., Mendigutia, I., Ilee, J.~D., \& van den Ancker, M.~E.\ 2017, \mnras, 464, 4721 

\bibitem[Fang et al.(2009)]{fang2009} Fang, M., van Boekel, R., Wang, W., et al.\ 2009, \aap, 504, 461 

\bibitem[Follette et al.(2017)]{follette2017} Follette, K.~B., Rameau, J., Dong, R., et al.\ 2017, \aj, 153, 264 

\bibitem[Furlan et al.(2006)]{furlan2006} Furlan, E., Hartmann, L., Calvet, N., et al.\ 2006, \apjs, 165, 568

\bibitem[Furlan et al.(2011)]{furlan2011} Furlan, E., Luhman, K.~L., Espaillat, C., et al.\ 2011, \apjs, 195, 3

\bibitem[Gaia Collaboration et al.(2018)]{GAIA2018} Gaia Collaboration, Brown, A.~G.~A., Vallenari, A., et al.\ 2018, arXiv:1804.09365

\bibitem[Gail(2001)]{gail2001} Gail, H.-P.\ 2001, \aap, 378, 192 

\bibitem[Garrison(1967)]{Garrison} Garrison, R.~F.\ 1967, \pasp, 79, 433

\bibitem[Garufi et al.(2016)]{garufi2016} Garufi, A., Quanz, S.~P., Schmid, H.~M., et al.\ 2016, \aap, 588, A8 

\bibitem[Geers et al.(2006)]{geers2006} Geers, V.~C., Augereau, J.-C., Pontoppidan, K.~M., et al.\ 2006, \aap, 459, 545 

\bibitem[Gullbring et al.(1998)]{gullbring1998} Gullbring, E., Hartmann, L., Brice{\~n}o, C., \& Calvet, N.\ 1998, \apj, 492, 323

\bibitem[Gutermuth et al.(2009)]{gutermuth2009} Gutermuth, R.~A., Megeath, S.~T., Myers, P.~C., et al.\ 2009, \apjs, 184, 18 

\bibitem[Habart et al.(2004)]{Natta04} Habart, E., Natta, A., \& Kr{\"u}gel, E.\ 2004, \aap, 427, 179

\bibitem[Harker \& Desch(2002)]{harker_desch2002} Harker, D.~E., \& Desch, S.~J.\ 2002, \apjl, 565, L109 

\bibitem[Henning \& Meeus(2011)]{henning2011} Henning, T., \& Meeus, G.\ 2011, Physical Processes in Circumstellar Disks around Young Stars, 114 

\bibitem[Hern{\'a}ndez et al.(2004)]{hernandez2004} Hern{\'a}ndez, J., Calvet, N., Brice{\~n}o, C., Hartmann, L., \& Berlind, P.\ 2004, \aj, 127, 1682

\bibitem[Hern{\'a}ndez et al.(2007)]{hernandez2007} Hern{\'a}ndez, J., Hartmann, L., Megeath, T., et al.\ 2007, \apj, 662, 1067 

\bibitem[Hern{\'a}ndez et al.(2009)]{hernandez2009} Hern{\'a}ndez, J., Calvet, N., Hartmann, L., et al.\ 2009, \apj, 707, 705

\bibitem[Hern{\'a}ndez et al.(2010)]{hernandez2010} Hern{\'a}ndez, J., Morales-Calderon, M., Calvet, N., et al.\ 2010, \apj, 722, 1226 

\bibitem[Hern{\'a}ndez et al.(2014)]{hernandez2014} Hern{\'a}ndez, J., Calvet, N., Perez, A., et al.\ 2014, \apj, 794, 36 

\bibitem[Higdon et al.(2004)]{smart2004} Higdon, S.~J.~U., Devost, D., Higdon, J.~L., et al.\ 2004, \pasp, 116, 975 

\bibitem[Houck et al.(2004)]{Houck04} Houck, J.~R., Roellig, T.~L., van Cleve, J., et al.\ 2004, \apjs, 154, 18 

\bibitem[Ingleby et al.(2013)]{ingleby2013} Ingleby, L., Calvet, N., Herczeg, G., et al.\ 2013, \apj, 767, 112

\bibitem[Davis(1999)]{iraf1999} Davis, L.~E.\ 1999, Precision CCD Photometry, 189, 35 

\bibitem[Ishihara et al.(2010)]{ishihara2010} Ishihara, D., Onaka, T., Kataza, H., et al.\ 2010, \aap, 514, A1

\bibitem[Jeffries et al.(2006)]{jeffries2006} Jeffries, R.~D., Maxted, P.~F.~L., Oliveira, J.~M., \& Naylor, T.\ 2006, \mnras, 371, L6 

\bibitem[Juh{\'a}sz et al.(2012)]{juhasz2012} Juh{\'a}sz, A., Dullemond, C.~P., van Boekel, R., et al.\ 2012, \apj, 744, 118 

\bibitem[Juh{\'a}sz et al.(2010)]{juhasz2010} Juh{\'a}sz, A., Bouwman, J., Henning, T., et al.\ 2010, \apj, 721, 431 

\bibitem[Kamp(2011)]{evolPAH} Kamp, I.\ 2011, EAS Publications Series, 46, 271 

\bibitem[Keller et al.(2008)]{Keller08} Keller, L.~D., Sloan, G.~C., Forrest, W.~J., et al.\ 2008, \apj, 684, 411-429 

\bibitem[Kenyon \& Hartmann(1995)]{kenyon1995} Kenyon, S.~J., \& Hartmann, L.\ 1995, \apjs, 101, 117 

\bibitem[Kessler-Silacci et al.(2006)]{kessler2006} Kessler-Silacci, J., Augereau, J.-C., Dullemond, C.~P., et al.\ 2006, \apj, 639, 275 

\bibitem[Kharchenko et al.(2007)]{kharchenko2007} Kharchenko, N.~V., Scholz, R.-D., Piskunov, A.~E., R{\"o}ser, S., \& Schilbach, E.\ 2007, Astronomische Nachrichten, 328, 889 

\bibitem[Kounkel et al.(2018)]{kounkel2018} Kounkel, M., Covey, K., Su{\'a}rez, G., et al.\ 2018, arXiv:1805.04649.

\bibitem[Lada et al.(2006)]{lada2006} Lada, C.~J., Muench, A.~A., Luhman, K.~L., et al.\ 2006, \aj, 131, 1574 

\bibitem[Landolt(2009)]{landolt2009} Landolt, A.~U.\ 2009, \aj, 137, 4186

\bibitem[Lebouteiller et al.(2010)]{smart} Lebouteiller, V., Bernard-Salas, J., Sloan, G.~C., \& Barry, D.~J.\ 2010, \pasp, 122, 231  

\bibitem[Lebouteiller et al.(2011)]{cassis} Lebouteiller, V., Barry, D.~J., Spoon, H.~W.~W., et al.\ 2011, \apjs, 196, 8

\bibitem[Li \& Draine(2002)]{li2002} Li, A., \& Draine, B.~T.\ 2002, \apj, 572, 232 

\bibitem[Luhman et al.(2010)]{luhman2010} Luhman, K.~L., Allen, P.~R., Espaillat, C., Hartmann, L., \& Calvet, N.\ 2010, \apjs, 186, 111

\bibitem[Maaskant et al.(2013)]{maaskant2013} Maaskant, K.~M., Honda, M., Waters, L.~B.~F.~M., et al.\ 2013, \aap, 555, A64 

\bibitem[Malfait et al.(1998)]{malfait1998} Malfait, K., Waelkens, C., Waters, L.~B.~F.~M., et al.\ 1998, \aap, 332, L25 

\bibitem[Massey \& Davis(1992)]{massey1992} Massey, P., \& Davis, L. E. 1992, A user's guide to stellar CCD photometry with IRAF, NOAO

\bibitem[Mathis et al.(1977)]{mathis1977} Mathis, J.~S., Rumpl, W., \& Nordsieck, K.~H.\ 1977, \apj, 217, 425 

\bibitem[Mathis(1990)]{mathis1990} Mathis, J.~S.\ 1990, \araa, 28, 37

\bibitem[Barrett et al.(2005)]{matplotlib2005} Barrett, P., Hunter, J., Miller, J.~T., Hsu, J.-C., \& Greenfield, P.\ 2005, Astronomical Data Analysis Software and Systems XIV, 347, 91 

\bibitem[Mattioda et al.(2005)]{mattioda2005} Mattioda, A.~L., Allamandola, L.~J., \& Hudgins, D.~M.\ 2005, \apj, 629, 1183 

\bibitem[Mauc{\'o} et al.(2018)]{mauco2018} Mauc{\'o}, K., Brice{\~n}o, C., Calvet, N., et al.\ 2018, \apj, 859, 1 

\bibitem[Mauc{\'o} et al.(2016)]{mauco2016} Mauc{\'o}, K., Hern{\'a}ndez, J., Calvet, N., et al.\ 2016, \apj, 829, 38

\bibitem[Maxted et al.(2008)]{maxted2008} Maxted, P.~F.~L., Jeffries, R.~D., Oliveira, J.~M., Naylor, T., \& Jackson, R.~J.\ 2008, \mnras, 385, 2210 

\bibitem[McClure et al.(2013)]{mcclure2013} McClure, M.~K., D'Alessio, P., Calvet, N., et al.\ 2013, \apj, 775, 114

\bibitem[McClure et al.(2010)]{mcclure2010} McClure, M.~K., Furlan, E., Manoj, P., et al.\ 2010, \apjs, 188, 75 

\bibitem[McClure et al.(2012)]{mcclure2012} McClure, M.~K., Manoj, P., Calvet, N., et al.\ 2012, \apjl, 759, L10 

\bibitem[Meeus et al.(2009)]{meeus2009} Meeus, G., Juh{\'a}sz, A., Henning, T., et al.\ 2009, \aap, 497, 379 

\bibitem[Meeus et al.(2001)]{meeus2001} Meeus, G., Waters, L.~B.~F.~M., Bouwman, J., et al.\ 2001, \aap, 365, 476 

\bibitem[Mulders et al.(2013)]{mulders2013} Mulders, G.~D., Paardekooper, S.-J., Pani{\'c}, O., et al.\ 2013, \aap, 557, A68 

\bibitem[Mulders et al.(2011)]{mulders2011} Mulders, G.~D., Waters, L.~B.~F.~M., Dominik, C., et al.\ 2011, \aap, 531, A93 

\bibitem[Muzerolle et al.(2003)]{muzerolle2003} Muzerolle, J., Calvet, N., Hartmann, L., \& D'Alessio, P.\ 2003, \apjl, 597, L149 

\bibitem[Muzerolle et al.(2010)]{muzerolle2010} Muzerolle, J., Allen, L.~E., Megeath, S.~T., Hern{\'a}ndez, J., \& Gutermuth, R.~A.\ 2010, \apj, 708, 1107 


\bibitem[Oliveira et al.(2006)]{oliveira2006} Oliveira, J., Bouwman, J., Jeffries, R., van Loon, J., \& van den Ancker, M.\ 2006, Spitzer Proposal, 30381 

\bibitem[Olofsson et al.(2009)]{olofsson2009} Olofsson, J., Augereau, J.-C., van Dishoeck, E.~F., et al.\ 2009, \aap, 507, 327

\bibitem[Owen(2016)]{owen2016} Owen, J.~E.\ 2016, \pasa, 33, e005

\bibitem[Pascual et al.(2016)]{pascual2016} Pascual, N., Montesinos, B., Meeus, G., et al.\ 2016, \aap, 586, A6 

\bibitem[Pecaut \& Mamajek(2013)]{pecaut2013} Pecaut, M.~J., \& Mamajek, E.~E.\ 2013, \apjs, 208, 9 

\bibitem[P{\'e}ricaud et al.(2017)]{pericaut2017} P{\'e}ricaud, J., Di Folco, E., Dutrey, A., Guilloteau, S., \& Pi{\'e}tu, V.\ 2017, \aap, 600, A62 

\bibitem[Pinilla et al.(2016)]{pinilla2016} Pinilla, P., Flock, M., Ovelar, M.~d.~J., \& Birnstiel, T.\ 2016, \aap, 596, A81 

\bibitem[Pinilla et al.(2015)]{pinilla2015} Pinilla, P., Birnstiel, T., \& Walsh, C.\ 2015, \aap, 580, A105 

\bibitem[Pollack et al.(1994)]{pollack1994} Pollack, J.~B., Hollenbach, D., Beckwith, S., et al.\ 1994, \apj, 421, 615 

\bibitem[Quanz et al.(2015)]{quanz2015} Quanz, S.~P., Amara, A., Meyer, M.~R., et al.\ 2015, \apj, 807, 64 

\bibitem[Rayner et al.(2009)]{rayner2009} Rayner, J.~T., Cushing, M.~C., \& Vacca, W.~D.\ 2009, \apjs, 185, 289 

\bibitem[Ribas et al.(2015)]{ribas2015} Ribas, {\'A}., Bouy, H., \& Mer{\'{\i}}n, B.\ 2015, \aap, 576, A52 

\bibitem[Roskosz et al.(2011)]{roskosz2011} Roskosz, M., Gillot, J., Capet, F., Roussel, P., \& Leroux, H.\ 2011, \aap, 529, A111 

\bibitem[Rubinstein et al.(2018)]{rubinstein2018} Rubinstein, A.~E., Macias, E., Espaillat, C.~C., et al.\ 2018, arXiv:1804.07343 

\bibitem[Ru{\'{\i}}z-Rodr{\'{\i}}guez et al.(2016)]{ruiz2016} Ru{\'{\i}}z-Rodr{\'{\i}}guez, D., Ireland, M., Cieza, L., \& Kraus, A.\ 2016, \mnras, 463, 3829 

\bibitem[Sacco et al.(2008)]{sacco2008} Sacco, G.~G., Franciosini, E., Randich, S., \& Pallavicini, R.\ 2008, \aap, 488, 167

\bibitem[Sargent et al.(2009a)]{sargent2009a} Sargent, B.~A., Forrest, W.~J., Tayrien, C., et al.\ 2009, \apj, 690, 1193

\bibitem[Sargent et al.(2009b)]{sargent2009b} Sargent, B.~A., Forrest, W.~J., Tayrien, C., et al.\ 2009, \apjs, 182, 477 

\bibitem[Sartori et al.(2010)]{sartori2010} Sartori, M.~J., Gregorio-Hetem, J., Rodrigues, C.~V., Hetem, A., Jr., \& Batalha, C.\ 2010, \aj, 139, 27-38

\bibitem[Schaefer(2013)]{schaefer2013} Schaefer, G.~H.\ 2013, EAS Publications Series, 64, 181

\bibitem[Seok \& Li(2017)]{seok2017} Seok, J.~Y., \& Li, A.\ 2017, \apj, 835, 291 

\bibitem[Shakura \& Sunyaev(1973)]{shakura1973} Shakura, N.~I., \& Sunyaev, R.~A.\ 1973, X- and Gamma-Ray Astronomy, 55, 155

\bibitem[Sicilia-Aguilar et al.(2013)]{sicilia2013} Sicilia-Aguilar, A., Kim, J.~S., Sobolev, A., et al.\ 2013, \aap, 559, A3 


\bibitem[Siess et al.(2000)]{siess2000} Siess, L., Dufour, E., \& Forestini, M.\ 2000, \aap, 358, 593 

\bibitem[Sivia \& Skilling (2012)]{sivia2012} Sivia, D., \& Skilling, \ 2012, Data Analysis: a Bayesian tutorial, 246 

\bibitem[Sim{\'o}n-D{\'{\i}}az et al.(2015)]{simon2015} Sim{\'o}n-D{\'{\i}}az, S., Caballero, J.~A., Lorenzo, J., et al.\ 2015, \apj, 799, 169 

\bibitem[Sloan et al.(2005)]{sloan2005} Sloan, G.~C., Keller, L.~D., Forrest, W.~J., et al.\ 2005, \apj, 632, 956 

\bibitem[Strom et al.(1989)]{strom1989} Strom, K.~M., Strom, S.~E., Edwards, S., Cabrit, S., \& Skrutskie, M.~F.\ 1989, \aj, 97, 1451

\bibitem[Sung et al.(2009)]{sung2009} Sung, H., Stauffer, J.~R., \& Bessell, M.~S.\ 2009, \aj, 138, 1116 

\bibitem[Testi et al.(2014)]{testi2014} Testi, L., Birnstiel, T., Ricci, L., et al.\ 2014, Protostars and Planets VI, 339

\bibitem[Torres et al.(1995)]{torres1995} Torres, C.~A.~O., Quast, G., de La Reza, R., Gregorio-Hetem, J., \& Lepine, J.~R.~D.\ 1995, \aj, 109, 2146 

\bibitem[van Boekel et al.(2005)]{vanBoekel2005} van Boekel, R., Min, M., Waters, L.~B.~F.~M., et al.\ 2005, \aap, 437, 189 

\bibitem[van Boekel et al.(2004)]{vanBoekel2004} van Boekel, R., Waters, L.~B.~F.~M., Dominik, C., et al.\ 2004, \aap, 418, 177 

\bibitem[van der Marel et al.(2016)]{vandermarel2016} van der Marel, N., Verhaar, B.~W., van Terwisga, S., et al.\ 2016, \aap, 592, A126 

\bibitem[Visser et al.(2007)]{Vpah07} Visser, R., Geers, V.~C., Dullemond, C.~P., et al.\ 2007, \aap, 466, 229 

\bibitem[Walsh et al.(2014)]{walsh2014} Walsh, C., Juh{\'a}sz, A., Pinilla, P., et al.\ 2014, \apjl, 791, L6 

\bibitem[Watson et al.(2009)]{watson2009} Watson, D.~M., Leisenring, J.~M., Furlan, E., et al.\ 2009, \apjs, 180, 84

\bibitem[White \& Basri(2003)]{white2003} White, R.~J., \& Basri, G.\ 2003, \apj, 582, 1109

\bibitem[Willacy et al.(2015)]{willacy2015} Willacy, K., Alexander, C., Ali-Dib, M., et al.\ 2015, \ssr, 197, 151 

\bibitem[Williams \& Cieza(2011)]{williams2011} Williams, J.P., \& Cieza, L.~A.\ 2011, \araa, 49, 67 

\bibitem[Williams et al.(2013)]{williams2013} Williams, J.~P., Cieza, L.~A., Andrews, S.~M., et al.\ 2013, \mnras, 435, 1671 

\bibitem[Wright et al.(2015)]{wright2015} Wright, C.~M., Maddison, S.~T., Wilner, D.~J., et al.\ 2015, \mnras, 453, 414 

\bibitem[Yamamura et al.(2010)]{yamamura2010} Yamamura, I., Makiuti, S., Ikeda, N., et al.\ 2010, VizieR Online Data Catalog, 2298

\bibitem[Zhang et al.(2015)]{zhang2015} Zhang, K., Blake, G.~A., \& Bergin, E.~A.\ 2015, \apjl, 806, L7 

\bibitem[Zhang et al.(2016)]{zhang2016} Zhang, K., Bergin, E.~A., Blake, G.~A., et al.\ 2016, \apjl, 818, L16 

\bibitem[Zhu et al.(2011)]{zhu2011} Zhu, Z., Nelson, R.~P., Hartmann, L., Espaillat, C., \& Calvet, N.\ 2011, \apj, 729, 47 

\bibitem[Zhu et al.(2012)]{zhu2012} Zhu, Z., Nelson, R.~P., Dong, R., Espaillat, C., \& Hartmann, L.\ 2012, \apj, 755, 6 
 
\end{thebibliography}
\end{document}